\documentclass[a4paper,12pt]{article}
\usepackage{amsmath}
\usepackage{amssymb}
\usepackage{revsymb}
\usepackage{mathrsfs}
\begin{document}
\title{Pair production in a strong magnetic field:
the effect of a weak background gravitational field}
\author{Antonino Di Piazza\footnote{E-mail:
dipiazza@ts.infn.it}
\\ \\
Dipartimento di Fisica Teorica, \\
Strada Costiera 11, Trieste, I-34014, Italy \\
 and \\
 INFN, Sezione di Trieste, Italy}
\maketitle
\begin{abstract}
The production probability of an $e^--e^+$ pair in the presence of a strong,
uniform and slowly varying magnetic field is calculated by taking into account
the presence of a background gravitational field. The curvature of the spacetime
metric induced by the gravitational field not only changes the transition
probabilities calculated in the Minkowski spacetime but also primes transitions
that are strictly forbidden in absence of the gravitational field.

\noindent PACS numbers: 04.62.+v, 12.20.Ds, 98.70.Rz
\end{abstract}
\clearpage
\section{Introduction}
In view of the persistent interest in the high energy emissions from astrophysical compact objects \cite{GRB}, we present a continuation of the study on the electron-positron pairs
production in the presence of strong ($\gg B_{cr}=m^2c^3/(\hbar e)
\simeq 4.4\times 10^{13}\; \mbox{gauss}$), time varying magnetic fields
\cite{Calucci,DiPiazza1,DiPiazza2}. As we have pointed out in our previous papers,
the physical situation we have in mind is the creation of $e^--e^+$ pairs
around astrophysical compact objects like neutron stars or black holes that can produce
such strong magnetic fields \cite{Thompson,Hanami}. In this picture, the time variation of the magnetic field is a consequence of the rotation of the compact object or its gravitational collapse following a supernova explosion. Also, the
creation of the electrons and of the positrons is only an intermediate step.
In fact, our final goal is to give a possible theoretical interpretation of
the so-called gamma-ray bursts (GRBs) that are thought to be originated as a consequence of a supernova explosion around such astrophysical compact objects \cite{GRB}.  In our formulation GRBs are pulses of photons produced by the electrons
and the positrons created by the time dependent stellar magnetic field through their annihilation \cite{DiPiazza3} and/or as synchrotron radiation \cite{DiPiazza4}. Nevertheless, we are aware that the real physical situation is more complicated than our theoretical model for various reasons. In fact, on the one hand we treat just one process of pair production while, actually, other processes (the production of $e^--e^+$ pairs by photons, see the review \cite{Erber}) are present and can give dominant contributions. On the other hand, we stress much more ``microscopic'' features of the pair (photon) production such as selection rules that allow or prevent some particular elementary processes and so on by neglecting ``macroscopic'' aspects that would be very difficult to be treated analytically (the macroscopic structure of the compact object magnetic field or the presence of already created particles). These aspects are, instead, privileged in other papers where the microscopic details are not brought out (see e.g. \cite{Kogan1,Kogan2,Biermann,Kogan3}). In these papers, inspired by the suggestion in \cite{Kardashev} that nonthermal radio emissions can be accounted by the gravitational collapse of a massive star, a similar interpretation is given about the origin of a GRB as a consequence of a supernova explosion. In turn, the supernova explosion is thought to be caused by a combination of gravitational and magnetic factors and much more emphasis is given to the order of magnitude of the energy released or to other macroscopic features of the physical situation.

Instead, till now, we have neglected the effects of the gravitational field created by the
neutron star or by the black hole by performing all the calculations in a flat spacetime. Even if there are situations in which this can be safely done \cite{Heyl,Gibbons1}, it is interesting to study what happens if the effects of the gravitational field are taken into
account. Actually, since the early work by Hawking on black holes
thermal emission \cite{Hawking}, it has been widely demonstrated from
a theoretical point of view that a gravitational field can be
responsible by itself of particle creation (see Refs. \cite{Stellare}
for particle creation in the presence of stellar gravitational fields
and Refs. \cite{Cosmologico} for cosmological particle creation). The
formalism used to deal with this subject is that of quantum field
theory in curved spacetimes where the gravitational field is treated
classically and the consequent curvature of the spacetime metric is
assigned, while the matter fields (Klein-Gordon or Dirac fields) are
quantized starting from a Lagrangian density written in a general
covariant way \cite{Birrell,Fulling,Wald}. In the same framework, the
possibility of particle creation in the presence of gravitational and
electromagnetic fields has also been investigated \cite{Gibbons2,Villalba}. 

In the cited papers the particle creation is due to the fact that the gravitational field is time-dependent and the Bogoliubov transformation technique is used to calculate the production rate. Instead, as we will explain in the following section, despite we start by using the same
formalism, then we change our point of view. In fact, in our model
the main process responsible for pair creation is still the magnetic field and
the fact that it varies with time. The gravitational field is static and it enters the
calculations because it modifies the one-particle states and energies
of the electrons and positrons and consequently the production probabilities.
In the present work, the structure of the gravitational field is assumed to be such that its
effects can be calculated perturbatively.

The paper is structured as follows. In section \ref{II} the physical assumptions and the consequent theoretical model is described: the gravitational effects are given starting from the Schwarzschild metric and then expanding it around a specific point, not too close to the event horizon and ending the expansion at the first derivative of the metric tensor. Then the magnetic field in this metric is given and the Hamiltonian of the second quantized electron field is built up. In section \ref{III} the one-particle eigenstates and eigenenergies of the Dirac field in the given magnetic field with the perturbative effect of the gravity are displayed. In section \ref{IV} the pair production due to the time dependence of the magnetic field (in the constant gravitational field) are calculated with two explicit form of  time variation. Some conclusions are presented in section \ref{V}. Finally, four appendices contain explicit calculations that would have made heavy the main text.

In what follows natural units ($\hbar=c=1$) are used throughout and the
signature of the Minkowski spacetime is assumed to be $+---$.
%
%
\section{Theoretical framework}
\label{II}
\setcounter{equation}{0}
\renewcommand{\theequation}{II.\arabic{equation}}

As we have said in the Introduction, we want to calculate the probability
of producing an electron-positron pair in the presence of a strong, time
varying magnetic field and of a gravitational field created by a neutron
star or by a black hole. Even if we assume that the spatial structure and
the time evolution of both fields are given, the problem formulated in
these terms is very difficult to be faced. In fact, to determine a
realistic form of the gravitational field and of the (electro)magnetic
field we should fix the physical properties of the source (its mass, its
electric charge, its angular momentum and so on) and solve the system
formed by the Einstein equations and the general covariant Maxwell
equations. Clearly, solving the Einstein equations coupled with the general covariant Maxwell
equations is a hopeless problem and a number of
approximations have to be done. In particular, we first assume that the
Einstein equations and the general covariant Maxwell equations are
disentangled. This corresponds to neglect the gravitational field produced
by the magnetic field and to assume that the spacetime metric is determined
only by the stellar object. But, in order to fit our requests this stellar
object should be capable to produce a time-varying magnetic field. In
this sense, it could be a charged rotating black hole or a magnetized
collapsing neutron star but the corresponding spacetime metrics would
be still very difficult to deal with. The problem can be definitely
simplified by assuming that the spacetime metric is actually that
produced by an uncharged, non rotating star and that the corrections to
this metric due to the charge or to the rotation of the star itself
can be neglected. In this approximations, our starting point is the
metric tensor corresponding to the field created by a spherical body
of mass $M$ outside the body itself. If we call $t$, $X$, $Y$ and $Z$
the four coordinates, this metric tensor can be written as \cite{Landau2}
\begin{equation}
\label{g_mu_nu}
g_{\mu\nu}(X,Y,Z)=\text{diag}\left[\frac{F^2_-(X,Y,Z)}{F^2_+(X,Y,Z)},
-F^4_+(X,Y,Z),
-F^4_+(X,Y,Z),-F^4_+(X,Y,Z)\right]
\end{equation}
where
\begin{equation}
\label{F_pm}
F_{\pm}(X,Y,Z)=1\pm\frac{r_g}{4\sqrt{X^2+Y^2+Z^2}}
\end{equation}
with $r_g=2GM$ the gravitational radius of the body ($G$ is the gravitational
constant). We have chosen the so-called isotropic metric instead of the usual
(and equivalent) Schwarzschild metric, because from Eq. (\ref{g_mu_nu}) we see
that the spatial distance is proportional to its Euclidean expression and this
will simplify our future calculations. We point out that in this metric the
event horizon of the body is the spherical surface $\sqrt{X^2+Y^2+Z^2}=r_g/4$.

Now, an $e^--e^+$ pair is created in a spatial volume with typical length
given by the Compton length $\lambdabar=1/m$ ($m$ is the electron mass). Since
$\lambdabar$ is much smaller than the gravitational radius $r_g$ of a neutron
star or of a black hole, we are allowed to make some simplifications on the
metric tensor (\ref{g_mu_nu}). In particular, we can always assume that the
pair is created in a small neighborhood of the space point $P_0$ labeled by
the coordinates $(X_0,0,0)$ with $X_0>r_g/4+\Delta$ and $\Delta>0$. If
$P=(X,Y,Z)=(X_0+x,y,z)$ is a generic point in this neighborhood then
$|x|\lesssim\lambdabar$, $|y|\lesssim\lambdabar$ and $|z|\lesssim\lambdabar$
and we can approximate the metric tensor $g_{\mu\nu}(P)$ calculated in $P$
with the metric tensor
\begin{equation}
g^{(1)}_{\mu\nu}(x,y,z)=g_{\mu\nu}(P_0)+\left.\frac{\partial g_{\mu\nu}(P)}
{\partial X}\right\vert_{P=P_0}x+\left.\frac{\partial g_{\mu\nu}(P)}{\partial Y}
\right\vert_{P=P_0}y+\left.\frac{\partial g_{\mu\nu}(P)}{\partial Z}\right\vert_{P=P_0}z
\end{equation}
in which only the terms up to first order in $x/X_0$, $y/X_0$ and $z/X_0$
have been kept. It can easily be seen that $g^{(1)}_{\mu\nu}(x,y,z)$ actually
depends only on $x$ and that it can be written in the form
\begin{equation}
\label{g_mu_nu_l}
g^{(1)}_{\mu\nu}(x)=g^{(0)}_{\mu\nu}+h_{\mu\nu}(x)
\end{equation}
where
\begin{align}
\label{g_mu_nu_l_2}
g^{(0)}_{\mu\nu} &\equiv g_{\mu\nu}(P_0)=\text{diag}(\phi_t,-\phi_s,-\phi_s,-\phi_s),\\
\label{g_mu_nu_l_2_h}
h_{\mu\nu}(x) &\equiv\left.\frac{\partial g_{\mu\nu}(P)}{\partial X}\right
\vert_{P=P_0}x=\text{diag}(2g_t x,2g_s x,2g_s x,2g_s x)
\end{align}
with [see Eqs. (\ref{g_mu_nu}) and (\ref{F_pm})]
\begin{align}
\label{Phi_c}
\phi_t &=\left(1-\frac{r_g}{4X_0}\right)^2\bigg/\left(1+\frac{r_g}{4X_0}\right)^2,
& \phi_s &=\left(1+\frac{r_g}{4X_0}\right)^4,\\
\label{g}
g_t &=\left(1-\frac{r_g}{4X_0}\right)\frac{r_g}{2X_0^2}\bigg/\left(1+\frac{r_g}
{4X_0}\right)^3, & g_s &=\left(1+\frac{r_g}{4X_0}\right)^3\frac{r_g}{2X_0^2}.
\end{align}

It is evident that, in order that $g^{(1)}_{\mu\nu}(x)$ is a good approximation
of $g_{\mu\nu}(P)$, $X_0$ can not be chosen to be too close to the critical
value $r_g/4$. Just to give an idea, it easy to see that, if $N\gg 1$ is a
large pure number, then
\begin{align}
\left\vert \frac{g_{\mu\mu}(P)-g^{(1)}_{\mu\mu}(x)}{g_{\mu\mu}(P)}\right
\vert<\frac{1}{N} && \text{no sum}
\end{align}
with $\mu=0,\ldots,3$, if
\begin{equation}
\label{ineq}
X_0>\frac{r_g}{4}+\sqrt{N}\lambdabar.
\end{equation}
This condition automatically implies that
\begin{align}
\left\vert \frac{h_{\mu\mu}(x)}{g^{(0)}_{\mu\mu}}\right\vert<\frac{1}{2\sqrt{N}}
&& \text{no sum}
\end{align}
with $\mu=0,\ldots,3$ and then that $h_{\mu\nu}(x)$ can be considered as a
small correction of $g^{(0)}_{\mu\nu}$. In what follows, we assume that the
previous inequalities hold with sufficiently large $N$.

Now, we pass to the description of the structure of the magnetic field that
we will denote as $\mathbf{B}(\mathbf{r},t)$ and that, for simplicity, we
will assume to lie in the $y-z$ plane: $\mathbf{B}(\mathbf{r},t)=
[0,B_y(\mathbf{r},t),B_z(\mathbf{r},t)]$. With similar arguments used in
the case of the gravitational field we can assume that the magnetic field
is uniform and slowly varying in time in the spacetime volume where an $e^--e^+$ pair is created. By writing it as
\begin{equation}
\label{B}
\mathbf{B}(\mathbf{r},t)=\mathbf{B}(t)=
\left(\begin{array}{c}
0\\
B_y(t)\\
B_z(t)
\end{array}\right)=B(t)\left(\begin{array}{c}
0\\
\sin\vartheta(t)\\
\cos\vartheta(t)
\end{array}\right)
\end{equation}
with
\begin{align}
B(t) &=\sqrt{B^2_y(t)+B^2_z(t)},\\
\tan\vartheta(t) &=\frac{B_y(t)}{B_z(t)}
\end{align}
and by also reminding that we work in the strong-field regime, the magnetic
field is such that
\begin{align}
\label{str_f}
B(t) &\gg B_{cr}=\frac{m^2}{e},\\
\label{sl_var}
\frac{\big\vert\dot{\mathbf{B}}(t)\big\vert }{B(t)} &\ll \frac{1}{\lambdabar}=m
\end{align}
where $-e<0$ is the electron charge. Before going on we want to make two
observations about the magnetic field. Firstly, we have shown that the magnetic
field (\ref{B}) does not satisfy the Maxwell equations in vacuum and in the
spacetime with the metric (\ref{g_mu_nu_l}), but an electric current is needed
to produce it. Secondly, we point out that the assumptions that the gravitational
field is static while $\mathbf{B}(t)$ is time varying are not contradictory
even if the two fields are produced by the same source. From a theoretical
point of view, this circumstance happens, for example, for a spherical body
which collapses keeping its spherical symmetry and without rotating. In this
case, in fact, the gravitational field of the body is static because of the
Birkhoff theorem \cite{Weinberg} while the magnetic field is found to be
time-dependent from energy-conservation considerations. Obviously, the
previous conclusion is correct if the gravitational field generated by the
magnetic energy can be neglected.

Since the spacetime is curved and the metric tensor is not simply $\eta_{\mu\nu}$,
we must pay attention in defining the vector potential $A_{\mu}(\mathbf{r},t)$
that gives rise to $\mathbf{B}(t)$. To this end, we assume that the
three-dimensional components of the magnetic field $\mathbf{B}(t)$ define
the spatial-spatial components of the full covariant electromagnetic tensor
$F_{\mu\nu}(\mathbf{r},t)$ that is
\begin{align}
\label{F_i_j_x}
F_{32}(t)=-F_{23}(t) &\equiv 0,\\
F_{13}(t)=-F_{31}(t) &\equiv B_y(t),\\
\label{F_i_j_z}
F_{21}(t)=-F_{12}(t) &\equiv B_z(t)
\end{align}
while the mixed or the full contravariant components are built by using the
metric tensor (\ref{g_mu_nu_l}). Now, by definition
\begin{equation}
\label{F_mu_nu}
F_{\mu\nu}(\mathbf{r},t)\equiv A_{\nu ;\mu}(\mathbf{r};t)-A_{\mu ;\nu}
(\mathbf{r},t)=A_{\nu ,\mu}(\mathbf{r},t)-A_{\mu ,\nu}(\mathbf{r},t),
\end{equation}
then by means of Eqs. (\ref{F_i_j_x})-(\ref{F_i_j_z}), we can choose the
covariant vector $A_{\mu}(\mathbf{r},t)$ as
\begin{align}
A_0(\mathbf{r},t) &=0,\\
A_1(\mathbf{r},t) &=\frac{1}{2}\left[\mathbf{r}\times\mathbf{B}(t)\right]_x,\\
A_2(\mathbf{r},t) &=\frac{1}{2}\left[\mathbf{r}\times\mathbf{B}(t)\right]_y,\\
A_3(\mathbf{r},t) &=\frac{1}{2}\left[\mathbf{r}\times\mathbf{B}(t)\right]_z.
\end{align}
These relations define a gauge analogous to the so-called ``symmetric gauge''
\cite{DiPiazza2} in the Minkowski spacetime. Finally, it is convenient to
define the three-dimensional vector potential $\mathbf{A}(\mathbf{r},t)=
[A_x(\mathbf{r},t),A_y(\mathbf{r},t),A_z(\mathbf{r},t)]$ as
\begin{equation}
\label{A}
\mathbf{A}(\mathbf{r},t)=-\frac{1}{2}\left[\mathbf{r}\times\mathbf{B}(t)\right]
\end{equation}
where the minus sign has been inserted to have $\boldsymbol{\partial}\times
\mathbf{A}(\mathbf{r},t)=\mathbf{B}(t)$.

As we have done in our previous papers \cite{Calucci,DiPiazza1,DiPiazza2},
we will calculate the probability to create the $e^--e^+$ pair from the
vacuum by applying the adiabatic perturbation theory up to first order in
the time derivative $\dot{\mathbf{B}}(t)$ of the magnetic field \cite{Migdal}.
To do this, we have to build up the second quantized Hamiltonian of a Dirac
field $\Psi (\mathbf{r},t)$ in the presence of the slowly varying magnetic
field (\ref{B}) and in the curved spacetime with the static metric tensor
(\ref{g_mu_nu_l}) and to determine its instantaneous eigenstates and
eigenenergies \cite{Migdal}. The Lagrangian density of this system is
given by \cite{Birrell}:
\begin{equation}
\label{L}
\begin{split}
\mathscr{L}(\Psi,\partial_{\mu}\Psi,\bar{\Psi},\partial_{\mu}\bar{\Psi},\mathbf{r},t)=
\sqrt{-g^{(1)}(x)}\bigg\{
&\frac{1}{2}\Big[\bar{\Psi}\gamma^{(1)\,\mu}(x)[i\partial_{\mu}+
i\Gamma_{\mu}(x)+eA_{\mu}(\mathbf{r},t)]\Psi-\\
&\left.-\bar{\Psi}[i\overleftarrow{\partial}_{\mu}-i\Gamma_{\mu}(x)-
eA_{\mu}(\mathbf{r},t)]\gamma^{(1)\,\mu}(x)\Psi\right]-m\bar{\Psi}\Psi\bigg\}.
\end{split}
\end{equation}
The presence of the gravitational field is represented in Eq. (\ref{L})
by the following quantities
\begin{align}
\label{g_l}
g^{(1)}(x) &\equiv\det[g^{(1)}_{\mu\nu}(x)],\\
\gamma^{(1)\,\mu}(x) &\equiv\gamma^{\alpha}e_{\alpha}^{(1)\mu}(x),\\
\label{nabla_mu}
\Gamma_{\mu}(x) &=-\frac{i}{4}\sigma^{\alpha\beta}e_{\alpha}^{(1)\nu}(x)
e_{\beta\nu;\mu}^{(1)}(x)
\end{align}
where $\gamma^{\alpha}$ are the usual Dirac matrices satisfying
$\{\gamma^{\alpha},\gamma^{\beta}\}=2\eta^{\alpha\beta}$ (assumed in the
Dirac representation), $e_{\alpha}^{(1)\mu}(x)$ is a tetrad corresponding
to the metric (\ref{g_mu_nu_l}) \cite{Weinberg,Birrell} and where
$\sigma^{\alpha\beta}=i[\gamma^{\alpha},\gamma^{\beta}]/2$. We remind that
in Eq. (\ref{L}) the adjoint field $\bar{\Psi}(\mathbf{r},t)$ is also
defined as $\bar{\Psi}(\mathbf{r},t)\equiv \Psi^{\dag}(\mathbf{r},t)\gamma^0$
in a curved spacetime.

Since $g^{(1)}_{\mu\nu}(x)$ has been split as in Eq. (\ref{g_mu_nu_l}) with
the matrix $h_{\mu\nu}(x)$ much smaller than the matrix $g^{(0)}_{\mu\nu}$,
we are allowed to keep in Eq. (\ref{L}) only the first-order terms in
$h_{\mu\nu}(x)$. To do this, we need to calculate the quantities
(\ref{g_l})-(\ref{nabla_mu}) up to first order. Concerning the determinant
$g^{(1)}(x)$, we have immediately
\begin{equation}
\label{g_l_expl}
g^{(1)}(x)\simeq g^{(0)}[1+h(x)]
\end{equation}
where
\begin{align}
\label{g_0}
g^{(0)} &\equiv\det(g^{(0)}_{\mu\nu})=-\phi_t\phi_s^3,\\
\label{h}
h(x) &\equiv h_{\mu}^{\;\mu}(x)=2\left(\frac{g_t}{\phi_t}-\frac{3g_s}{\phi_s}\right)x.
\end{align}
Also, being the metric tensor $g^{(1)}_{\mu\nu}(x)$ diagonal, we can
choose a diagonal tetrad with
\begin{align}
\label{V_0^(1)1}
e_0^{(1)0}(x) &=\frac{1}{\sqrt{g^{(1)}_{00}(x)}}\simeq\frac{1}{\sqrt{\phi_t}}
\left(1-\frac{g_tx}{\phi_t}\right),\\
e_i^{(1)i}(x) &=\frac{1}{\sqrt{-g^{(1)}_{ii}(x)}}\simeq\frac{1}{\sqrt{\phi_s}}
\left(1+\frac{g_sx}{\phi_s}\right) && \text{no sum}
\end{align}
(note that while the Greek indices run from zero to three, the Latin ones
run from one to three). By means of this tetrad it can be shown that the
connections $\Gamma_{\mu}(x)$ are already first-order quantities equal to
\begin{equation}
\Gamma_{\mu}^{(1)}(x)=\frac{i}{4}\sigma^{1\beta}e_{1}^{(0)1}e_{\beta}^{(0)\rho}
\frac{dh_{\mu\rho}(x)}{dx}
\end{equation}
where $e_{\alpha}^{(0)\mu}$ is the diagonal zero-order tetrad with
\begin{align}
e_0^{(0)0} &=\frac{1}{\sqrt{g^{(0)}_{00}}}=\frac{1}{\sqrt{\phi_t}},\\
e_i^{(0)i} &=\frac{1}{\sqrt{-g^{(0)}_{ii}}}=\frac{1}{\sqrt{\phi_s}}  && \text{no sum}.
\end{align}

Now, the procedure to calculate the Lagrangian density (\ref{L}) up to first
order in $h_{\mu\nu}(x)$ is identical to that used in the weak-field
approximation \cite{Gupta,Ogievetskii} and we give only its final expression:
\begin{equation}
\label{L_l}
\begin{split}
\mathscr{L}^{(1)}(\Psi,\partial_{\mu}\Psi,\bar{\Psi},\partial_{\mu}\bar{\Psi},
\mathbf{r},t)=\sqrt{\phi_t\phi_s^3} \bigg\{ &\frac{(1-g_Ex)}{2\sqrt{\phi_t}}
[\bar{\Psi}\gamma^0(i\partial_0\Psi)-(i\partial_0\bar{\Psi})\gamma^0\Psi]+\\
&+\frac{(1-g_Px)}{2\sqrt{\phi_s}}\Big[\bar{\Psi}\gamma^i[i\partial_i+
eA_i(\mathbf{r},t)]\Psi-\\
&\qquad\qquad\qquad-\bar{\Psi}[i\overleftarrow{\partial}_i-eA_i(\mathbf{r},t)]
\gamma^i\Psi\Big]-\\
&-(1-g_Mx)m\bar{\Psi}\Psi\bigg\}
\end{split}
\end{equation}
where we defined the three couplings\footnote{We introduced three couplings
for later convenience because, actually, only two of them are independent.}
\begin{align}
\label{g_E}
g_E &\equiv\frac{3g_s}{\phi_s}=\frac{3r_g}{2X_0^2}\bigg/\left(1+\frac{r_g}{4X_0}\right),\\
\label{g_P}
g_P &\equiv\frac{2g_s}{\phi_s}-\frac{g_t}{\phi_t}=\frac{r_g}{2X_0^2}
\left[2-\left(1-\frac{r_g}{4X_0}\right)^{-1}\right]\bigg/\left(1+\frac{r_g}{4X_0}\right),\\
\label{g_M}
g_M &\equiv\frac{3g_s}{\phi_s}-\frac{g_t}{\phi_t}=\frac{r_g}{2X_0^2}
\left[3-\left(1-\frac{r_g}{4X_0}\right)^{-1}\right]\bigg/\left(1+\frac{r_g}{4X_0}\right).
\end{align}
Note that the modifications induced by the metric tensor (\ref{g_mu_nu_l})
in the Lagrangian density (\ref{L_l}) are linear in $g_t$ and $g_s$ (obviously)
but nonlinear in $\phi_t$ and $\phi_s$.

The definition of the Hamiltonian is a controversial operation in a
curved spacetime \cite{Fulling_p_1,Grib,Castagnino}. We shall adopt the
same definition given in \cite{Fulling_p_2} for a scalar field that is
\begin{equation}
\label{H_def}
\mathscr{H}^{(1)}(\Psi,\partial_i\Psi,\bar{\Psi},\partial_i\bar{\Psi},\Pi^{(1)},
\bar{\Pi}^{(1)},\mathbf{r},t)\equiv \bar{\Pi}^{(1)}(\partial_0\Psi)+(\partial_0
\bar{\Psi})\Pi^{(1)}-\mathscr{L}^{(1)}(\Psi,\partial_{\mu}\Psi,\bar{\Psi},
\partial_{\mu}\bar{\Psi},\mathbf{r},t)
\end{equation}
where
\begin{align}
\bar{\Pi}^{(1)}(\mathbf{r},t) &\equiv
\frac{\partial \mathscr{L}^{(1)}}{\partial (\partial_0\Psi)},\\
\Pi^{(1)}(\mathbf{r},t) &\equiv\frac{\partial \mathscr{L}^{(1)}}{\partial (\partial_0\bar{\Psi})}
\end{align}
are the first-order conjugated momenta to the fields $\Psi(\mathbf{r},t)$ and
$\bar{\Psi}(\mathbf{r},t)$ respectively. In our case,
\begin{align}
\bar{\Pi}^{(1)}(\mathbf{r},t) &=\sqrt{\phi_s^3}\frac{i(1-g_Ex)}{2}
\bar{\Psi}(\mathbf{r},t)\gamma^0,\\
\Pi^{(1)}(\mathbf{r},t) &=-\sqrt{\phi_s^3}\frac{i(1-g_Ex)}{2}\gamma^0\Psi(\mathbf{r},t)
\end{align}
and then the Hamiltonian density (\ref{H_def}) becomes
\begin{equation}
\label{H}
\begin{split}
\mathscr{H}^{(1)}(\Psi,\partial_i\Psi,\bar{\Psi},\partial_i\bar{\Psi},\mathbf{r},t)
&=\sqrt{\phi_t\phi_s^3}\Big\{-\frac{(1-g_Px)}{2\sqrt{\phi_s}}\Big[\bar{\Psi}
\gamma^i[i\partial_i+eA_i(\mathbf{r},t)]\Psi-\\
&\qquad\qquad\qquad\qquad\qquad\quad-\bar{\Psi}[i\overleftarrow{\partial}_i-
eA_i(\mathbf{r},t)]\gamma^i\Psi\Big]+\\
&\qquad\qquad\quad+(1-g_Mx)m\bar{\Psi}\Psi\bigg\}
\end{split}
\end{equation}
and it results independent of $\Pi^{(1)}(\mathbf{r},t)$ and $\bar{\Pi}^{(1)}
(\mathbf{r},t)$. Coherently, the total Hamiltonian is defined as \cite{Fulling_p_2}
\begin{equation}
\label{H_tot_l}
H^{(1)}(t)\equiv\int d\mathbf{r}\mathscr{H}^{(1)}(\Psi,\partial_i\Psi,\bar{\Psi},\mathbf{r},t).
\end{equation}
We point out that this definition can be shown to be equivalent in our case to
the alternative definition which uses the energy-momentum tensor
(see \cite{Fulling_p_1} for a detailed discussion about the relation between
these two definitions).
Now,
\begin{equation}
(1-g_Px)(\partial_i\bar{\Psi})\gamma^i\Psi=\partial_i[(1-g_Px)\bar{\Psi}\gamma^i\Psi]-
\bar{\Psi}\gamma^i\partial_i[(1-g_Px)\Psi]
\end{equation}
then the Hamiltonian density (\ref{H}) is equivalent apart from a derivative
term to the asymmetric Hamiltonian density (we use the same symbol to indicate it)
\begin{equation}
\label{H_a}
\begin{split}
\mathscr{H}^{(1)}(\Psi,\partial_i\Psi,\bar{\Psi},\mathbf{r},t)=\sqrt{\phi_t\phi_s^3}
\bigg\{ &-\frac{1}{2\sqrt{\phi_s}}\Big[(1-g_Px)\bar{\Psi}\gamma^i[i\partial_i+
eA_i(\mathbf{r},t)]\Psi+\\
&\qquad\qquad+\bar{\Psi}\gamma^i[i\partial_i+eA_i(\mathbf{r},t)][(1-g_Px)\Psi]\Big]+\\
&+(1-g_Mx)m\bar{\Psi}\Psi\bigg\}.
\end{split}
\end{equation}
This Hamiltonian density has the advantage that it can be written in the form
\begin{equation}
\label{H_a_1}
\mathscr{H}^{(1)}(\Psi,\partial_i\Psi,\bar{\Psi},\mathbf{r},t)=\sqrt{\phi_s^3}
(1-g_Ex)\Psi^{\dag}\mathcal{H}^{(1)}(\mathbf{r},-i\boldsymbol{\partial},t)\Psi
\end{equation}
where $\Psi^{\dag}(\mathbf{r},t)=\bar{\Psi}(\mathbf{r},t)\gamma^0$ and where we
introduced the one-particle first-order Hamiltonian
\begin{equation}
\label{H_1p}
\begin{split}
\mathcal{H}^{(1)}(\mathbf{r},-i\boldsymbol{\partial},t) &=\frac{\sqrt{\phi_t}}
{1-g_Ex}\bigg\{\frac{1}{2\sqrt{\phi_s}}\Big[(1-g_Px)\boldsymbol{\alpha}
\cdot[-i\boldsymbol{\partial}+e\mathbf{A}(\mathbf{r},t)]+\\
&\qquad\qquad\quad\;+\boldsymbol{\alpha}\cdot[-i\boldsymbol{\partial}+
e\mathbf{A}(\mathbf{r},t)](1-g_Px)\Big]+(1-g_Mx)\beta m\bigg\}\simeq\\
&\simeq \sqrt{\phi_t}\bigg\{\frac{1}{2\sqrt{\phi_s}}\Big[(1-g_Px)\boldsymbol{\alpha}
\cdot[-i\boldsymbol{\partial}+e\mathbf{A}(\mathbf{r},t)]+\\
&\qquad\qquad +\boldsymbol{\alpha}\cdot[-i\boldsymbol{\partial}+
e\mathbf{A}(\mathbf{r},t)](1-g_Px)\Big]+(1-g_Mx)\beta m+\\
&\qquad\qquad +g_Ex\left[\frac{1}{\sqrt{\phi_s}}\boldsymbol{\alpha}
\cdot[-i\boldsymbol{\partial}+e\mathbf{A}(\mathbf{r},t)]+\beta m\right]\bigg\}
\end{split}
\end{equation}
with $\beta=\gamma^0$ and $\boldsymbol{\alpha}=(\alpha_x,\alpha_y,\alpha_z)=
(\beta\gamma^1,\beta\gamma^2,\beta\gamma^3)$. Before explaining the reason
of this apparently unusual definition we observe that the one-particle
Hamiltonian (\ref{H_1p}) can be written as the sum
\begin{equation}
\label{H_1p_2}
\mathcal{H}^{(1)}(\mathbf{r},-i\boldsymbol{\partial},t)=
\mathcal{H}^{(0)}(\mathbf{r},-i\boldsymbol{\partial},t)+
\mathcal{I}(\mathbf{r},-i\boldsymbol{\partial},t)
\end{equation}
of the zero-order Hamiltonian
\begin{equation}
\label{H_1p_0}
\mathcal{H}^{(0)}(\mathbf{r},-i\boldsymbol{\partial},t)=\sqrt{\frac{\phi_t}{\phi_s}}
\left\{\boldsymbol{\alpha}\cdot[-i\boldsymbol{\partial}+e\mathbf{A}(\mathbf{r},t)]+
\sqrt{\phi_s}\beta m\right\}
\end{equation}
and of the first-order interaction
\begin{equation}
\label{U_1p}
\begin{split}
\mathcal{I}(\mathbf{r},-i\boldsymbol{\partial},t) &=\sqrt{\phi_t}
\bigg\{ -\frac{g_P}{2\sqrt{\phi_s}}\Big[x\boldsymbol{\alpha}
\cdot[-i\boldsymbol{\partial}+e\mathbf{A}(\mathbf{r},t)]+\boldsymbol{\alpha}
\cdot[-i\boldsymbol{\partial}+e\mathbf{A}(\mathbf{r},t)]x\Big]-g_M \beta mx+\\
&\qquad\qquad\qquad\qquad+g_Ex\left[\frac{1}{\sqrt{\phi_s}}\boldsymbol{\alpha}
\cdot[-i\boldsymbol{\partial}+e\mathbf{A}(\mathbf{r},t)]+\beta m\right]\bigg\}=\\
&=\sqrt{\phi_t}(g_P-g_M)\beta mx-\frac{g_P}{2}
\left[x\mathcal{H}^{(0)}(\mathbf{r},-i\boldsymbol{\partial},t)+
\mathcal{H}^{(0)}(\mathbf{r},-i\boldsymbol{\partial},t)x\right]+\\
&\quad+g_Ex\mathcal{H}^{(0)}(\mathbf{r},-i\boldsymbol{\partial},t).
\end{split}
\end{equation}

In order to understand the definition (\ref{H_1p}) we have to introduce
the scalar product for spinors in our curved spacetime. If
$\zeta_1(\mathbf{r},t)$ and $\zeta_2(\mathbf{r},t)$ are two spinors
it is defined as
\begin{equation}
\label{s_p}
(\zeta_1,\zeta_2)\equiv\int_{S}dS_{\mu}\sqrt{-g^{(1)}(x)}
\bar{\zeta}_1(\mathbf{r},t)\gamma^{(1)\mu}(x)\zeta_2(\mathbf{r},t)
\end{equation}
where $S$ is a Cauchy hyper-surface. We can choose $S$ as the $t=\text{const.}$
hyper-surface, then \cite{Landau2}
\begin{equation}
dS_{\mu}=\left(d\mathbf{r},0,0,0\right)
\end{equation}
and the scalar product (\ref{s_p}) becomes
\begin{equation}
(\zeta_1,\zeta_2)=\int d\mathbf{r}\sqrt{-g^{(1)}(x)}\zeta_1^{\dag}
(\mathbf{r},t)e_0^{(1)0}(x)\zeta_2(\mathbf{r},t)=\int d\mathbf{r}
\sqrt{-\frac{g^{(1)}(x)}{g^{(1)}_{00}(x)}}\zeta_1^{\dag}(\mathbf{r},t)\zeta_2(\mathbf{r},t)
\end{equation}
where we used the definition (\ref{V_0^(1)1}). Finally, by exploiting
Eqs. (\ref{g_l_expl})-(\ref{h}), the volume element can be written up to first order as
\begin{equation}
\label{d_sigma}
d\mathbf{r}\sqrt{-\frac{g^{(1)}(x)}{g^{(1)}_{00}(x)}}\simeq d\mathbf{r}
\sqrt{\frac{\phi_t\phi_s^3}{\phi_t}}\left(1+\frac{g_t x}{\phi_t}-
\frac{3g_s x}{\phi_s}-\frac{g_t x}{\phi_t}\right)=d\mathbf{r}\sqrt{\phi_s^3}(1-g_Ex)
\end{equation}
and we obtain
\begin{equation}
\label{s_p_f}
(\zeta_1,\zeta_2)=\int d\mathbf{r}\sqrt{\phi_s^3}(1-g_Ex)\zeta_1^{\dag}
(\mathbf{r},t)\zeta_2(\mathbf{r},t).
\end{equation}
Now, with this definition of the scalar product we realize that the
``asymmetric'' Hamiltonian (\ref{H_1p}) is actually an Hermitian operator
up to first-order terms. The fact that this definition of one-particle
Hamiltonian is well posed is also corroborated by the form of the equation
of motion of the field $\Psi(\mathbf{r},t)$. In fact, starting from the
Lagrangian density (\ref{L_l}), it can be seen that the equation of motion
\begin{equation}
\frac{\partial}{\partial t}\frac{\partial \mathscr{L}^{(1)}}{\partial
(\partial_0\bar{\Psi})}+\boldsymbol{\partial}\cdot\frac{\partial
\mathscr{L}^{(1)}}{\partial (\boldsymbol{\partial}\bar{\Psi})}=0
\end{equation}
can be written as
\begin{equation}
\label{Eq_m}
i\partial_0\Psi=\mathcal{H}^{(1)}(\mathbf{r},-i\boldsymbol{\partial},t)\Psi.
\end{equation}
Finally, by using Eq. (\ref{H_a_1}), the total Hamiltonian (\ref{H_tot_l}) becomes
\begin{equation}
\label{H_tot_f}
H^{(1)}(t)=\int d\mathbf{r}\sqrt{\phi_s^3}(1-g_Ex)\Psi^{\dag}(\mathbf{r},t)
\mathcal{H}^{(1)}(\mathbf{r},-i\boldsymbol{\partial},t)\Psi(\mathbf{r},t).
\end{equation}
This Hamiltonian depends explicitly on time only through the dependence of
$\mathcal{H}^{(1)}(\mathbf{r},-i\boldsymbol{\partial},t)$ on the magnetic
field $\mathbf{B}(t)$, then because of the condition (\ref{sl_var}) it is a
slowly varying quantity. In the next section we will determine its
instantaneous eigenstates and eigenvalues that will be used to compute
the pair creation probability by means of the adiabatic perturbation
theory \cite{Migdal}.
%
%
\section{Determination of the instantaneous eigenstates and eigenenergies
of the Hamiltonian $H^{(1)}(t)$}
\label{III}
\setcounter{equation}{0}
\renewcommand{\theequation}{III.\arabic{equation}}

The first step to determine the instantaneous eigenstates and eigenenergies
of the Hamiltonian (\ref{H_tot_f}) is to quantize and diagonalize the
corresponding constant Hamiltonian. If we introduce the constant magnetic field
\begin{equation}
\label{B_ti}
\mathbf{B}=
\left(\begin{array}{c}
0\\
B_y\\
B_z
\end{array}\right)=B\left(\begin{array}{c}
0\\
\sin\vartheta\\
\cos\vartheta
\end{array}\right)
\end{equation}
with
\begin{align}
\label{B_mod_ti}
B &=\sqrt{B^2_y+B^2_z},\\
\label{theta_ti}
\tan\vartheta &=\frac{B_y}{B_z}
\end{align}
and the vector potential in the symmetric gauge
\begin{equation}
\label{A_ti}
\mathbf{A}(\mathbf{r})=-\frac{1}{2}(\mathbf{r}\times\mathbf{B})
\end{equation}
then the Lagrangian density (\ref{L_l}), the one-particle Hamiltonian (\ref{H_1p_2})
and the total Hamiltonian (\ref{H_tot_f}) become respectively
\begin{align}
\label{L_l_ti}
\begin{split}
\mathscr{L}^{(1)}(\Psi,\partial_{\mu}\Psi,\bar{\Psi},\partial_{\mu}\bar{\Psi},\mathbf{r})
&=\sqrt{\phi_t\phi_s^3} \bigg\{\frac{(1-g_Ex)}{2\sqrt{\phi_t}}[\bar{\Psi}
\gamma^0(i\partial_0\Psi)-(i\partial_0\bar{\Psi})\gamma^0\Psi]+\\
&\qquad\qquad\quad+\frac{(1-g_Px)}{2\sqrt{\phi_s}}\Big[\bar{\Psi}\gamma^i
[i\partial_i+eA_i(\mathbf{r})]\Psi-\\
&\qquad\qquad\qquad\qquad\qquad\quad-\bar{\Psi}[i\overleftarrow{\partial}_i-
eA_i(\mathbf{r})]\gamma^i\Psi\Big]-\\
&\qquad\qquad\quad-(1-g_Mx)m\bar{\Psi}\Psi\bigg\},
\end{split}\\
\label{H_1p_2_ti}
\mathcal{H}^{(1)}(\mathbf{r},-i\boldsymbol{\partial}) &=\mathcal{H}^{(0)}
(\mathbf{r},-i\boldsymbol{\partial})+\mathcal{I}(\mathbf{r},-i\boldsymbol{\partial}),
\\
\label{H_tot_f_ti}
H^{(1)} &=\int d\mathbf{r}\sqrt{\phi_s^3}(1-g_Ex)\Psi^{\dag}(\mathbf{r},t)
\mathcal{H}^{(1)}(\mathbf{r},-i\boldsymbol{\partial})\Psi(\mathbf{r},t)
\end{align}
where [see Eqs. (\ref{H_1p_0}) and (\ref{U_1p})]
\begin{align}
\label{H_1p_0_ti}
\mathcal{H}^{(0)}(\mathbf{r},-i\boldsymbol{\partial}) &=\sqrt{\frac{\phi_t}{\phi_s}}
\left\{\boldsymbol{\alpha}\cdot[-i\boldsymbol{\partial}+e\mathbf{A}(\mathbf{r})]+
\sqrt{\phi_s}\beta m\right\},\\
\label{U_1p_ti}
\begin{split}
\mathcal{I}(\mathbf{r},-i\boldsymbol{\partial}) &=\sqrt{\phi_t}(g_P-g_M)\beta m x-
\frac{g_P}{2}\left[x\mathcal{H}^{(0)}(\mathbf{r},-i\boldsymbol{\partial})+
\mathcal{H}^{(0)}(\mathbf{r},-i\boldsymbol{\partial})x\right]+\\
&\quad+g_Ex\mathcal{H}^{(0)}(\mathbf{r},-i\boldsymbol{\partial}).
\end{split}
\end{align}
Also, the equation of motion (\ref{Eq_m}) becomes
\begin{equation}
\label{Eq_m_ti}
i\partial_0\Psi=\mathcal{H}^{(1)}(\mathbf{r},-i\boldsymbol{\partial})\Psi
\end{equation}
then, if
\begin{equation}
\Psi(\mathbf{r},t)=\Psi_{\pm,\jmath}(\mathbf{r})\exp(\mp iw_{\jmath}t)
\end{equation}
with $\jmath$ a set of quantum numbers and $\Psi_{\pm,\jmath}(\mathbf{r})$ are a complete set of modes with energies $w_{\jmath} >0$, then
\begin{equation}
\label{Eq_eig}
\mathcal{H}^{(1)}(\mathbf{r},-i\boldsymbol{\partial})\Psi_{\pm,\jmath}=
\pm w_{\jmath}\Psi_{\pm,\jmath}.
\end{equation}
After applying the charge conjugation operator to the positive-energy solution
$\tilde{\Psi}_{+,\jmath}(\mathbf{r})$ of the equation (\ref{Eq_eig}) with
$e$ instead of $-e$ in the one-particle Hamiltonian $\mathcal{H}^{(1)}
(\mathbf{r},-i\boldsymbol{\partial})$ and calling the corresponding spinor
$V_{\jmath}(\mathbf{r})$ (see \cite{DiPiazza2} for details), Eq. (\ref{Eq_eig})
splits into two equations:
\begin{align}
\label{Eq_eig_u}
\mathcal{H}^{(1)}(\mathbf{r},-i\boldsymbol{\partial})U_{\jmath} &=w_{\jmath} U_{\jmath},\\
\label{Eq_eig_v}
\mathcal{H}^{(1)}(\mathbf{r},-i\boldsymbol{\partial})V_{\jmath} &=-\tilde{w}_{\jmath} V_{\jmath}
\end{align}
where $U_{\jmath}(\mathbf{r})\equiv\Psi_{+,\jmath}(\mathbf{r})$ and where
we called $\tilde{w}_{\jmath}>0$ the energies of the positron states that,
in the presence of a magnetic field, have, in general, a different
dependence on the quantum numbers $\jmath$ \cite{DiPiazza2}. The states
$U_{\jmath}(\mathbf{r})$ and $V_{\jmath}(\mathbf{r})$ are assumed to be an
orthonormal basis with respect to the scalar product (\ref{s_p_f}), \textit{i.e.}
\begin{align}
\label{norm_1}
(U_{\jmath},U_{\jmath'}) &=(V_{\jmath},V_{\jmath'})=\delta_{\jmath,\jmath'},\\
\label{norm_2}
(U_{\jmath},V_{\jmath'}) &=(V_{\jmath},U_{\jmath'})=0.
\end{align}

Equations (\ref{Eq_eig_u}) and (\ref{Eq_eig_v}) with the orthonormalization
relations (\ref{norm_1}) and (\ref{norm_2}) will be solved perturbatively in
the next paragraph. At the moment, we observe that, in order to quantize and
diagonalize the Hamiltonian (\ref{H_tot_f_ti}), we have to expand the Dirac
field $\Psi(\mathbf{r},t)$ in the basis $[U_{\jmath}(\mathbf{r}),V_{\jmath}(\mathbf{r})]$ as
\begin{equation}
\Psi (\mathbf{r},t)=\sum_{\jmath} \left[c_{\jmath}(t) U_{\jmath}(\mathbf{r})+
d_{\jmath}^{\dag}(t) V_{\jmath} (\mathbf{r})\right]
\end{equation}
with $c_{\jmath}(t)=c_{\jmath} \exp(-iw_{\jmath}t)$ and $d_{\jmath}^{\dag}(t)=
d_{\jmath}^{\dag} \exp(i\tilde{w}_{\jmath}t)$
and we have to impose the usual anti-commutation rules among the coefficients
$c_{\jmath}$, $c_{\jmath}^{\dag}$, $d_{\jmath}$ and $d_{\jmath}^{\dag}$ that
are now operators. In this way, by using the equation of motion (\ref{Eq_m_ti}),
it can easily be shown that the Hamiltonian (\ref{H_tot_f_ti}) assumes the usual
diagonal form
\begin{equation}
\label{H_q}
H^{(1)}=\int d\mathbf{r}\sqrt{\phi_s^3}(1-g_Ex)\Psi^{\dag}(\mathbf{r},t)
\mathcal{H}^{(1)}(\mathbf{r},-i\boldsymbol{\partial})\Psi(\mathbf{r},t)=
\sum_{\jmath} (w_{\jmath}N_{\jmath}+\tilde{w}_{\jmath}\tilde{N}_{\jmath})
\end{equation}
where $N_{\jmath}=c_{\jmath}^{\dag}c_{\jmath}$, $\tilde{N}_{\jmath}=
d_{\jmath}^{\dag}d_{\jmath}$ and where we neglected the vacuum energy.
\footnote{Note that to neglect the vacuum energy can be unsafe when we want
to determine the time evolution of the gravitational field in the presence of
quantum matter fields as its sources \cite{Birrell}. But in our case this fact
does not cause any problems because the time evolution of the gravitational
field is given (actually, in the present case the gravitational field is static).} Obviously, if $|0\rangle$ is the vacuum state the eigenstates
of this Hamiltonian are the Fock states
\begin{equation}
\label{n_n_t}
|\{n_{\jmath}\};\{\tilde{n}_{\tilde{\jmath}}\}\rangle\equiv
\left(c^{(1)\dag}_{\jmath_1}\right)^{n_{\jmath_1}}
\left(c^{(1)\dag}_{\jmath_2}\right)^{n_{\jmath_2}}\cdots
\left(d^{(1)\dag}_{\tilde{\jmath}_1}\right)^{\tilde{n}_{\tilde{\jmath}_1}}
\left(d^{(1)\dag}_{\tilde{\jmath}_2}\right)^{\tilde{n}_{\tilde{\jmath}_2}}\cdots|0\rangle
\end{equation}
with the corresponding eigenvalues
\begin{equation}
\label{E}
E=\sum_l (w_{\jmath_l}n_{\jmath_l}+\tilde{w}_{\tilde{\jmath}_l}\tilde{n}_{\tilde{\jmath}_l}).
\end{equation}

In the next paragraph, we proceed by explicitly computing the one-particle
spinors $U_{\jmath}(\mathbf{r})$ and $V_{\jmath}(\mathbf{r})$ and the
energies $w_{\jmath}$ and $\tilde{w}_{\jmath}$ and we will come back to
the problem of the instantaneous eigenstates and eigenvalues of the
Hamiltonian (\ref{H_tot_f}) in the paragraph (\ref{III_3}). By looking at
the structure of the one-particle Hamiltonian (\ref{H_1p_2_ti}), we see that
the eigenvalue equations (\ref{Eq_eig_u}) and (\ref{Eq_eig_v}) can be solved
perturbatively with respect to the interaction Hamiltonian
$\mathcal{I}(\mathbf{r},-i\boldsymbol{\partial})$ by using the time-independent
perturbation theory \cite{Landau3}. In particular, we will compute the
eigenstates $U_{\jmath}(\mathbf{r})$ and $V_{\jmath}(\mathbf{r})$ and the
eigenenergies $w_{\jmath}$ and $\tilde{w}_{\jmath}$ up to first order in
the gravitational couplings $g_E$, $g_P$ and $g_M$.
%
%
\subsection{Computation of the one-particle states $U_{\jmath}(\mathbf{r})$ and
$V_{\jmath}(\mathbf{r})$ up to zero order and of the energies $w_{\jmath}$ and
$\tilde{w}_{\jmath}$ up to first order}

We first want to determine the zero-order solution of Eqs. (\ref{Eq_eig_u}) and
(\ref{Eq_eig_v}). Up to this order those equations can be written as
\begin{align}
\label{Eq_eig_u_0}
\sqrt{\frac{\phi_t}{\phi_s}}\left\{\boldsymbol{\alpha}\cdot[-i\boldsymbol{\partial}+
e\mathbf{A}(\mathbf{r})]+\sqrt{\phi_s}\beta m\right\}U_{\jmath}^{(0)} &=
w_{\jmath}^{(0)} U_{\jmath}^{(0)},\\
\label{Eq_eig_v_0}
\sqrt{\frac{\phi_t}{\phi_s}}\left\{\boldsymbol{\alpha}\cdot[-i\boldsymbol{\partial}+
e\mathbf{A}(\mathbf{r})]+\sqrt{\phi_s}\beta m\right\}V_{\jmath}^{(0)} &=
-\tilde{w}_{\jmath}^{(0)} V_{\jmath}^{(0)}.
\end{align}
These equations are the eigenvalue equations of a particle with charge $-e$ and
mass $\sqrt{\phi_s}m$ in the presence of the magnetic field (\ref{B_ti}) in the
Minkowski spacetime, then we can write immediately their solutions
\cite{Cohen,Bagrov}. We have already studied these solutions in \cite{DiPiazza2}
and we refer the reader to that paper for a detailed discussion about the
meaning of the quantum numbers that will be introduced, their ranges of
variation and so on.\footnote{We warn the reader that some unavoidable changes
of notation have been done with respect to Ref. \cite{DiPiazza2}.} Here, we
only quote the results that will be useful later. Firstly, we remind that a
complete set of commuting observables of this physical system is made by the
Hamiltonian $\mathcal{H}^{(0)}(\mathbf{r},-i\boldsymbol{\partial})$, the
longitudinal linear momentum $\mathcal{P}_{\parallel}$, the longitudinal total
angular momentum $\mathcal{J}_{\parallel}=\mathcal{L}_{\parallel}+
\mathcal{S}_{\parallel}$ and the transverse square distance $R^2_{\perp}$ where
``longitudinal'' and ``transverse'' are considered with respect to the
direction of the magnetic field. For this reason, the eigenstates and the
eigenenergies are characterized by four quantum numbers that are
$j\equiv\{n_d,k,\sigma,n_g\}$. Actually, the eigenenergies of the electron
do not depend on $n_g$ and those of the positron do not depend on $n_d$
[see Eqs. (15) and (34) in Ref. \cite{DiPiazza2}]. This circumstance leads
us to embody in $r$ the quantum numbers $\{n_d,k,\sigma\}$ and in $q$ the
quantum numbers $\{n_g,k,\sigma\}$. In conclusion, we have
\begin{align}
j &\equiv\{n_d,k,\sigma,n_g\},\\
\label{r}
r &\equiv\{n_d,k,\sigma\},\\
\label{q}
q &\equiv\{n_g,k,\sigma\}
\end{align}
and, consequently
\begin{equation}
j=\{r,n_g\}=\{q,n_d\}.
\end{equation}
With these definitions the electron and positron eigenenergies $w_j^{(0)}$ and
$\tilde{w}_j^{(0)}$ are given by the modified Landau levels
\begin{align}
\label{w_0}
w_r^{(0)} &=\sqrt{\frac{\phi_t}{\phi_s}}\sqrt{\phi_s m^2+k^2+eB(2n_d+1+\sigma)}=
\sqrt{\phi_t m^2+\frac{\phi_t}{\phi_s}\left[k^2+eB(2n_d+1+\sigma)\right]},\\
\label{w_t_0}
\tilde{w}_q^{(0)} &=\sqrt{\frac{\phi_t}{\phi_s}}\sqrt{\phi_s m^2+k^2+
eB(2n_g+1-\sigma)}=\sqrt{\phi_t m^2+\frac{\phi_t}{\phi_s}\left[k^2+eB(2n_g+1-\sigma)\right]}.
\end{align}
The corresponding eigenstates will be indicated as $u_j(\mathbf{r})$ and
$v_j(\mathbf{r})$ respectively and, since we have already studied them in
\cite{DiPiazza2}, we will quote some of their properties in appendix A.

As it is evident from the expressions (\ref{w_0}) and (\ref{w_t_0}), the
one-particle eigenenergies have two kinds of degenerations. The first one
is due, as we have said, to the fact that the electron eigenenergies do
not depend on the quantum number $n_g$ and, symmetrically, the positron
eigenenergies do not depend on the quantum number $n_d$. The second one
is due to the fact that the electron eigenstates with quantum numbers
$\{r_+,n_g\}=\{n_d,k,+1,n_g\}$ and $\{r_-,n_g\}=\{n_d+1,k,-1,n_g\}$ have
the same energy whatever $n_g$ and, symmetrically, the positron eigenstates
with quantum numbers $\{\tilde{q}_+,n_d\}=\{n_d,k,+1,n_g+1\}$ and
$\{\tilde{q}_-,n_d\}=\{n_d,k,-1,n_g\}$ have the same energy whatever
$n_d$. This means, following the time-independent perturbation theory
\cite{Landau3}, that the eigenstates $u_j(\mathbf{r})$ and $v_j(\mathbf{r})$
will not represent, in general, the correct zero-order eigenfunctions
$U_{\jmath}^{(0)}(\mathbf{r})$ and $V_{\jmath}^{(0)}(\mathbf{r})$. For this
reason we indicated them by means of the symbols $u_j(\mathbf{r})$ and
$v_j(\mathbf{r})$.

Now, we will compute explicitly only the zero-order electron eigenstates and the
first-order electron eigenenergies, while the analogous results for the positron
eigenstates and eigenenergies will be only quoted. Following the time-independent
perturbation theory for degenerate states, we write the zero-order solutions of
Eq. (\ref{Eq_eig_u}) with a given energy as linear combinations of all the
degenerate eigenstates $u_j(\mathbf{r})$ and $v_j(\mathbf{r})$ with that
energy. We will characterize the new states by means of the index $x_0$ and
this choice will be understood later:
\begin{align}
\label{U_0^0}
U^{(0)}_{r_0,x_0}(\mathbf{r}) &=\sum_{n_g} \mathsf{P}^{(0)}_{r_0,x_0;r_0,n_g}
u_{r_0,n_g}(\mathbf{r}),\\
\label{U_m^0}
U^{(0)}_{r_-,x_0}(\mathbf{r}) &=\sum_{n_g} \left[\mathsf{P}^{(0)}_{r_-,x_0;r_-,n_g}
u_{r_-,n_g}(\mathbf{r})+\mathsf{P}^{(0)}_{r_-,x_0;r_+,n_g}u_{r_+,n_g}(\mathbf{r})\right],\\
\label{U_p^0}
U^{(0)}_{r_+,x_0}(\mathbf{r}) &=\sum_{n_g} \left[\mathsf{P}^{(0)}_{r_+,x_0;r_-,n_g}
u_{r_-,n_g}(\mathbf{r})+\mathsf{P}^{(0)}_{r_+,x_0;r_+,n_g}u_{r_+,n_g}(\mathbf{r})\right].
\end{align}
In the first of these equations we considered that the eigenenergies of the
so-called electron transverse ground states \cite{DiPiazza2}, characterized
by the quantum numbers $\{r_0,n_g\}$ with $r_0=\{0,k,-1\}$, are degenerate
only with respect to $n_g$. Note that in Eqs. (\ref{U_m^0}) and (\ref{U_p^0})
the resulting spinors have been characterized, with an abuse of notation, by
the quantum numbers $r_-$ and $r_+$ respectively, even if, in general, they
will be only eigenstate of the linear longitudinal momentum $\mathcal{P}_{\parallel}$
with eigenvalue $k$. But, the coefficients $\mathsf{P}^{(0)}_{r_0,x_0;r_0,n_g}$,
$\mathsf{P}^{(0)}_{r_{\pm},x_0;r_{\pm},n_g}$ and $\mathsf{P}^{(0)}_{r_{\mp},x_0;r_{\pm},n_g}$,
are the solutions of the secular equations \cite{Landau3}
\begin{align}
\label{Sec_eq_0}
&\sum_{n'_g}\left(\mathcal{I}_{r_0,n_g;r_0,n'_g}- \epsilon_{r_0,x_0}\delta_{n_g,n'_g}\right)
\mathsf{P}^{(0)}_{r_0,x_0;r_0,n'_g}=0,\\
\label{Sec_eq_m}
&\sum_{n'_g}\left[\left(\mathcal{I}_{r_-,n_g;r_-,n'_g}-\epsilon_{r_-,x_0}
\delta_{n_g,n'_g}\right)\mathsf{P}^{(0)}_{r_-,x_0;r_-,n'_g}+
\mathcal{I}_{r_-,n_g;r_+,n'_g}\mathsf{P}^{(0)}_{r_-,x_0;r_+,n'_g}\right]=0,\\
&\sum_{n'_g}\left[\left(\mathcal{I}_{r_+,n_g;r_+,n'_g}-\epsilon_{r_-,x_0}
\delta_{n_g,n'_g}\right)\mathsf{P}^{(0)}_{r_-,x_0;r_+,n'_g}+
\mathcal{I}_{r_+,n_g;r_-,n'_g}\mathsf{P}^{(0)}_{r_-,x_0;r_-,n'_g}\right]=0,\\
\label{Sec_eq_p}
&\sum_{n'_g}\left[\left(\mathcal{I}_{r_-,n_g;r_-,n'_g}-\epsilon_{r_+,x_0}
\delta_{n_g,n'_g}\right)\mathsf{P}^{(0)}_{r_+,x_0;r_-,n'_g}+
\mathcal{I}_{r_-,n_g;r_+,n'_g}\mathsf{P}^{(0)}_{r_+,x_0;r_+,n'_g}\right]=0,\\
\label{Sec_eq_p_2}
&\sum_{n'_g}\left[\left(\mathcal{I}_{r_+,n_g;r_+,n'_g}-\epsilon_{r_+,x_0}
\delta_{n_g,n'_g}\right)\mathsf{P}^{(0)}_{r_+,x_0;r_+,n'_g}+
\mathcal{I}_{r_+,n_g;r_-,n'_g}\mathsf{P}^{(0)}_{r_+,x_0;r_-,n'_g}\right]=0
\end{align}
where, in general,
\begin{equation}
\label{I_jjp}
\mathcal{I}_{jj'}\equiv\int d\mathbf{r}\sqrt{\phi_s^3} u^{\dag}_j(\mathbf{r})
\mathcal{I}(\mathbf{r},-i\boldsymbol{\partial})u_{j'}(\mathbf{r})
\end{equation}
and where $\epsilon_{r_0,x_0}$, $\epsilon_{r_{\pm},x_0}$ are the first-order
corrections (to be determined) to the eigenenergy of the eigenstates
$U^{(0)}_{r_0,x_0}(\mathbf{r})$ and $U^{(0)}_{r_{\pm},x_0}(\mathbf{r})$
respectively. In appendix B we show that the matrix elements
$\mathcal{I}_{r_{\pm},n_g;r_{\mp},n'_g}$ vanish.\footnote{Note that the
matrix elements $\mathcal{I}_{r_-,n_g;r_+,n'_g}$ and $\mathcal{I}_{r_+,n'_g;r_-,n_g}$
are not complex conjugated numbers because the operator $\mathcal{I}(\mathbf{r},
-i\boldsymbol{\partial})$ is not Hermitian with respect to the scalar product
implicitly used in the definition (\ref{I_jjp}).} This means that the perturbation
$\mathcal{I}(\mathbf{r},-i\boldsymbol{\partial})$ does not remove the energy
degeneracy of the eigenstates characterized by $\{r_-,n_g\}$ and $\{r_+,n_g\}$
and that all the coefficients $\mathsf{P}^{(0)}_{r_{\pm},x_0;r_{\mp},n_g}$ can
be put equal to zero. In other words, we have to diagonalize the perturbation
$\mathcal{I}(\mathbf{r},-i\boldsymbol{\partial})$ inside every subspace labeled
by the quantum numbers $r=\{n_d,k,\sigma\}$, and the zero-order eigenstates
are indeed characterized by the quantum numbers $r$. Also, equations (\ref{U_0^0})-
(\ref{U_p^0}) become simply
\begin{equation}
\label{U_0}
U^{(0)}_{r,x_0}(\mathbf{r})=\sum_{n_g} \mathsf{P}^{(0)}_{r,x_0;r,n_g}u_{r,n_g}(\mathbf{r})
\end{equation}
while equations (\ref{Sec_eq_0})-(\ref{Sec_eq_p_2}) can be written as the single equation
\begin{equation}
\label{Sec_eq}
\sum_{n'_g}\left(\mathcal{I}_{r,n_g;r,n'_g}- \epsilon_{r,x_0}\delta_{n_g,n'_g}\right)
\mathsf{P}^{(0)}_{r,x_0;r,n'_g}=0.
\end{equation}
Now, by means of the same technique used in appendix B we can show that the
matrix elements $\mathcal{I}_{r,n_g;r,n'_g}$ can be written as
\begin{equation}
\label{m_e}
\mathcal{I}_{r,n_g;r,n'_g}=\left[(g_P-g_M) \frac{\phi_tm^2}{w^{(0)}_r}+(g_E-g_P)
w^{(0)}_r\right]\int dx dy \theta^{\prime *}_{n_d,n_g}(x,y)x_0\theta'_{n_d,n'_g}(x,y)
\end{equation}
where the operator $x_0$ and the functions $\theta'_{n_d,n_g}(x,y)$ have been
defined in Eqs. (\ref{xy_perp}) and (\ref{phi_t}) respectively. Now, from Eqs.
(\ref{xi_0}), (\ref{xi_pr_0}) and (\ref{phi}) we see that only the transverse
functions $\theta'_{n_d,n_g}(x,y)$ in $u_j(\mathbf{r})$ depend on $n_g$, then
we have to determine the coefficients $\mathsf{P}^{(0)}_{r,x_0;r,n_g}$ in such
a way that the linear combination $\sum_{n_g} \mathsf{P}^{(0)}_{r,x_0;r,n_g}
\theta'_{n_d,n_g}(x,y)$ diagonalizes the operator $x_0$.\footnote{Now, it is
clear why we called the additional index $x_0$.} This linear combination is
given in \cite{Cohen} [it is Eq. (104) in Complement $\text{E}_{\text{VI}}$].
The coefficients $\mathsf{P}^{(0)}_{r,x_0;r,n_g}$ result independent of the
quantum numbers $r$ and are given by
\begin{equation}
\mathsf{P}^{(0)}_{r,x_0;r,n_g}=\mathsf{P}^{(0)}_{x_0,n_g}=\sqrt[4]{\frac{eB}{\pi}}
\frac{1}{\sqrt{2^{n_g} n_g!}}\mathrm{H}_{n_g}(\sqrt{eB}x_0)\exp\left(-\frac{eBx_0^2}{2}\right)
\end{equation}
where
\begin{equation}
\mathrm{H}_{n_g}(\sqrt{eB}x_0)=\frac{2^{n_g}}{\sqrt{\pi}}\int_{-\infty}^{\infty}
ds(\sqrt{eB}x_0+is)^{n_g}\exp\left(-s^2\right)
\end{equation}
is the $n_g$th-order Hermite polynomial \cite{Ryzhik}. In this way the spinors
$U^{(0)}_{r,x_0}(\mathbf{r})$ have the same form of the spinors $u_j(\mathbf{r})$,
but with the function $\theta'_{n_d,n_g}(x,y)$ substituted by the function
\begin{equation}
\Theta'_{n_d,x_0}(x,y)\equiv\sum_{n_g}\mathsf{P}^{(0)}_{x_0,n_g}\theta'_{n_d,n_g}(x,y).
\end{equation}
By using the expression (\ref{phi_t_f}) of $\theta'_{n_d,n_g}(x,y)$ we have
\begin{equation}
\label{phi_t_xg}
\begin{split}
\Theta'_{n_d,x_0}(x,y) &=\frac{1}{\sqrt{n_d!}}(a_d^{\dag})^{n_d}\sqrt[4]{\frac{eB}{\pi}}
\sqrt{\frac{eB}{2}}\exp\left[-\frac{eB}{2}\left(x_0^2+\frac{x^2+y^2}{2}\right)\right]
\frac{1}{\pi}\times\\
&\times\int_{-\infty}^{\infty}ds\exp\left(-s^2\right)\sum_{n_g}\frac{1}{n_g!}
\left[(\sqrt{eB}x_0+is)\sqrt{eB}(x-iy)\right]^{n_g}=\\
&=\frac{1}{\sqrt{n_d!}}(a_d^{\dag})^{n_d}\sqrt[4]{\frac{eB}{\pi}}\sqrt{\frac{eB}{2}}
\exp\left[-\frac{eB}{2}\left(x_0^2+\frac{x^2+y^2}{2}\right)\right]\frac{1}{\pi}\times\\
&\times  \exp\left[eBx_0(x-iy)\right]\int_{-\infty}^{\infty}ds
\exp\left[-s^2+i\sqrt{eB}(x-iy)s\right]=\\
&=\frac{1}{\sqrt{n_d!}}(a_d^{\dag})^{n_d}\sqrt[4]{\frac{eB}{\pi}}
\sqrt{\frac{eB}{2\pi}}\exp\left\{-\frac{eB}{2}\left[(x-x_0)^2-iy(x-2x_0)\right]\right\}
\end{split}
\end{equation}
where we used the formula \cite{Ryzhik}
\begin{align}
\int_{-\infty}^{\infty}ds\exp\left(-b_1^2s^2\pm b_2s\right)=\frac{\sqrt{\pi}}{b_1}
\exp\left(\frac{b_2^2}{4b_1^2}\right) && \text{Re}(b_1^2) >0.
\end{align}

From now on, the quantum number $n_g$ completely disappears in our calculations
because it is essentially substituted by the quantum number $x_0$. This circumstance
allows us to simplify the notation. In fact, we can eliminate the subscript $d$
from the remaining quantum number $n_d$ and from the related operators, that is
[see Eqs. (\ref{a_d}) and (\ref{a_d_c})]
\begin{align}
\label{sub_1}
a_d &\longrightarrow a\equiv \frac{1}{2}\left[\sqrt{\frac{eB}{2}}(x-iy)+
\sqrt{\frac{2}{eB}}\left(\frac{\partial}{\partial x}+
\frac{1}{i}\frac{\partial}{\partial y}\right)\right],
\\
\label{sub_2}
a_d^{\dag} &\longrightarrow a^{\dag}\equiv\frac{1}{2}\left[\sqrt{\frac{eB}{2}}(x+iy)-
\sqrt{\frac{2}{eB}}\left(\frac{\partial}{\partial x}-\frac{1}{i}\frac{\partial}
{\partial y}\right)\right],
\\
\label{sub_3}
n_d &\longrightarrow n.
\end{align}
Also, later calculations will be simplified if we discretize the eigenvalues $x_0$.
This can be done by imposing in the functions $\Theta'_{n_d,x_0}(x,y)$ the
periodicity condition at $x=0$
\begin{equation}
\exp\left(i\frac{eBx_0Y}{2}\right)=\exp\left(-i\frac{eBx_0Y}{2}\right)
\end{equation}
where $Y$ is the length of the quantization volume in the $y$-direction. In this
way, the allowed eigenvalues are given by
\begin{align}
x_{0;\ell}=\frac{2\ell\pi}{eBY} && \ell=0,\pm 1,\dots.
\end{align}
We point out that if we imposed the periodicity condition at $x_0\neq 0$ the
allowed eigenvalues would change. Nevertheless, since at the end of the
calculations we will perform the continuum limit $Y\to\infty$, we are not
interested in the exact values of the allowed eigenvalues but only in the
eigenstate density $\varrho (x_0)$ which is
\begin{equation}
\label{varrho}
\varrho (x_0)\equiv \frac{d\ell}{d x_0}=\frac{eBY}{2\pi}
\end{equation}
independently of $x_0$. For notational simplicity, we will still indicate
the discrete eigenvalues $x_{0;\ell}$ as $x_0$, then, taking into account
the substitutions (\ref{sub_1})-(\ref{sub_3}), the function (\ref{phi_t_xg}) becomes
\begin{equation}
\label{phi_t_xg_d}
\Theta'_{n,x_0}(x,y)=\frac{1}{\sqrt{n!}}(a^{\dag})^n\sqrt[4]{\frac{eB}{\pi Y^2}}
\exp\left\{-\frac{eB}{2}\left[(x-x_0)^2-iy(x-2x_0)\right]\right\}.
\end{equation}
The numerical factors have been chosen in such a way that if we define [analogously
to the two-dimensional spinors $\varphi'_j(\mathbf{r})$ given in Eq. (\ref{phi})]
the two-dimensional spinors
\begin{equation}
\label{Phi}
\Phi'_J(\mathbf{r})\equiv\frac{\exp(ikz)}{\sqrt{Z}}f'_{\sigma}\Theta'_{n,x_0}(x,y)
\end{equation}
with
\begin{equation}
J\equiv\{n,k,\sigma,x_0\},
\end{equation}
they result normalized as [see also Eq. (\ref{ort_phi_1})]
\begin{equation}
\label{ort_Phi}
\int d\mathbf{r}\Phi^{\prime\dag}_J(\mathbf{r})\Phi'_{J'}(\mathbf{r})=\delta_{J,J'}
\end{equation}
with $\delta_{J,J'}\equiv\delta_{n,n'}\delta_{k,k'}\delta_{\sigma,\sigma'}
\delta_{x_0,x_0'}$. Finally, the zero order spinors $U^{(0)}_J(\mathbf{r})$
are given by [see Eqs. (\ref{xi_0}) and (\ref{xi_pr_0})]
\begin{equation}
\label{U^0_f}
U^{(0)}_J(\mathbf{r})=\mathcal{R}^{\dag}_x(\vartheta)U^{\prime (0)}_J(\mathbf{r})
\end{equation}
where
\begin{equation}
\label{U^P0_f}
U^{\prime (0)}_J(\mathbf{r})= \frac{1}{\sqrt[4]{\phi_s^3}}\sqrt{\frac{w^{(0)}_R+
\sqrt{\phi_t}m}{2w^{(0)}_R}}
\left(\begin{array}{c}
      \Phi'_J(\mathbf{r})\\ \sqrt{\dfrac{\phi_t}{\phi_s}}
\dfrac{\mathcal{V}'(\mathbf{r},-i\boldsymbol{\partial})}{w_R^{(0)}+
\sqrt{\phi_t}m}\Phi'_J(\mathbf{r})
      \end{array}\right)
\end{equation}
with
\begin{equation}
R\equiv\{n,k,\sigma\}
\end{equation}
and are normalized as
\begin{equation}
\int d\mathbf{r}\sqrt{\phi_s^3}U^{(0)\dag}_J(\mathbf{r})U^{(0)}_{J'}(\mathbf{r})=\delta_{J,J'}.
\end{equation}
Obviously, from Eq. (\ref{m_e}) the eigenvalue corresponding to this
eigenfunction is, up to first order,
\begin{equation}
\label{w_1}
\begin{split}
w_J^{(1)} &=w_R^{(0)}+\left[(g_P-g_M) \frac{\phi_tm^2}{w^{(0)}_R}+
(g_E-g_P)w^{(0)}_R\right]x_0=\\
&=\sqrt{\phi_t m^2+\frac{\phi_t}{\phi_s}\left[k^2+eB(2n+1+\sigma)\right]}
\left\{1+\left[\frac{g_t}{\phi_t}+\frac{g_s}{\phi_s}\frac{k^2+eB(2n+1+\sigma)}
{\phi_sm^2+k^2+eB(2n+1+\sigma)}\right]x_0\right\}
\end{split}
\end{equation}
where we used the definitions (\ref{g_E})-(\ref{g_M}) of the coefficients
$g_E$, $g_P$ and $g_M$. This result has a clear interpretation in classical terms.
In fact, the classical Lagrangian of a particle in the given metric (\ref{g_mu_nu_l})
is \cite{Landau2}
\begin{align}
L(\mathbf{r},\mathbf{v})=-m\sqrt{g^{(1)}_{00}(x)+g^{(1)}_{ii}(x)v^2} && \text{no sum}
\end{align}
with
\begin{equation}
v^2=\left(\frac{dx}{dt}\right)^2+\left(\frac{dy}{dt}\right)^2+\left(\frac{dz}{dt}\right)^2
\end{equation}
or, up to first order,
\begin{equation}
L^{(1)}(\mathbf{r},\mathbf{v})=-m\sqrt{\phi_t-\phi_sv^2}
\left(1+\frac{g_t+g_sv^2}{\phi_t-\phi_sv^2}x\right).
\end{equation}

From this expression and by defining the first-order gravitational force on
the particle as \cite{Landau2}
\begin{equation}
\label{F_g}
\mathbf{f}^{(1)}(\mathbf{r},\mathbf{v})\equiv\frac{\partial L^{(1)}(\mathbf{r},\mathbf{v})}
{\partial \mathbf{r}}
\end{equation}
we easily obtain that the gravitational force lies along the $x$-axis and its $x$-component
is equal to
\begin{equation}
\label{f_1}
f^{(1)}_x(v)=-\frac{m\left(g_t+g_sv^2\right)}{\sqrt{\phi_t-\phi_sv^2}}.
\end{equation}
Finally, if we express $v^2$ in terms of the momentum $\mathbf{p}$ defined as
\begin{align}
\mathbf{p}\equiv\frac{\partial L(\mathbf{r},\mathbf{v})}{\partial \mathbf{v}}=
-\frac{mg^{(1)}_{ii}(x)\mathbf{v}}{\sqrt{g^{(1)}_{00}(x)+g^{(1)}_{ii}(x)v^2}}
&& \text{no sum},
\end{align}
we have
\begin{equation}
\label{F_g_x}
f^{(1)}_x(p)=f^{(1)}_x[v(p)]=-\sqrt{\phi_tm^2+\frac{\phi_t}{\phi_s}p^2}
\left(\frac{g_t}{\phi_t}+\frac{g_s}{\phi_s}\frac{p^2}{\phi_sm^2+p^2}\right)
\end{equation}
where, for simplicity, we used the same symbol to indicate the force as a
function of $v$ and of $p$. By performing the obvious identification
\begin{equation}
p^2\sim k^2+eB(2n+1+\sigma)
\end{equation}
and by comparing Eqs. (\ref{F_g_x}) and (\ref{w_1}), we conclude that the
first-order correction to the eigenenergies is nothing but the gravitational
potential energy of the electron in the presence of the constant and uniform
gravitational force (\ref{F_g_x}).

In an analogous way we can write the positron eigenenergies up to first order as
\begin{equation}
\label{w_t_1}
\tilde{w}_J^{(1)}=\sqrt{\phi_t m^2+\frac{\phi_t}{\phi_s}
\left[k^2+eB(2n+1-\sigma)\right]}\left\{1+\left[\frac{g_t}{\phi_t}+
\frac{g_s}{\phi_s}\frac{k^2+eB(2n+1-\sigma)}{\phi_sm^2+k^2+eB(2n+1-\sigma)}\right]x_0\right\}.
\end{equation}
We observe that, analogously to the case of the electron eigenstates and
eigenenergies, the quantum number $n_d$ is here substituted by $x_0$, and
it is pointless to keep the subscript $g$ in the remaining quantum number
$n_g$ because there is no more possibility of ambiguities.

The zero-order positron eigenstates are given by [see Eqs. (\ref{eta_0}) and (\ref{eta_pr_0})]
\begin{equation}
\label{V^0_f}
V^{(0)}_J(\mathbf{r})=\mathcal{R}^{\dag}_x(\vartheta)V^{\prime (0)}_J(\mathbf{r})
\end{equation}
with
\begin{equation}
\label{V^P0_f}
V^{\prime (0)}_J(\mathbf{r})= \frac{\sigma}{\sqrt[4]{\phi_s^3}}
\sqrt{\frac{\tilde{w}^{(0)}_R+\sqrt{\phi_t}m}{2\tilde{w}^{(0)}_R}}
\left(\begin{array}{c}
-\sqrt{\dfrac{\phi_t}{\phi_s}}\dfrac{\mathcal{V}'(\mathbf{r},
-i\boldsymbol{\partial})}{\tilde{w}_R^{(0)}+\sqrt{\phi_t}m}\mathrm{X}'_{J}(\mathbf{r})\\
      \mathrm{X}'_J(\mathbf{r})
      \end{array}\right)
\end{equation}
and [see Eq. (\ref{chi})]
\begin{equation}
\label{Chi}
\mathrm{X}'_J(\mathbf{r})\equiv\frac{\exp(-ikz)}{\sqrt{Z}}f'_{-\sigma}\Theta'_{n,x_0}(x,y).
\end{equation}
Obviously, these states are normalized as
\begin{equation}
\label{norm_V^0}
\int d\mathbf{r}\sqrt{\phi_s^3}V^{(0)\dag}_J(\mathbf{r})V^{(0)}_{J'}(\mathbf{r})=\delta_{J,J'}
\end{equation}
and are orthogonal to the eigenstates $U^{(0)}_J(\mathbf{r})$ up to zero-order terms:
\begin{equation}
\int d\mathbf{r}\sqrt{\phi_s^3}V^{(0)\dag}_J(\mathbf{r})U^{(0)}_{J'}(\mathbf{r})=
\int d\mathbf{r}\sqrt{\phi_s^3}U^{(0)\dag}_J(\mathbf{r})V^{(0)}_{J'}(\mathbf{r})=0.
\end{equation}
%
%
%
\subsection{Computation of the one-particle states $U_{\jmath}(\mathbf{r})$
and $V_{\jmath}(\mathbf{r})$ up to first order}

We can pass now to the determination of the one-particle states
$U_{\jmath}(\mathbf{r})$ and $V_{\jmath}(\mathbf{r})$ up to first order in the
perturbation $\mathcal{I}(\mathbf{r},-i\boldsymbol{\partial})$. From Eqs.
(\ref{w_1}) and (\ref{w_t_1}) we see that the first-order energies of the
transverse ground states that are given by
\begin{equation}
\label{w_g_1}
\varepsilon^{(1)}_{k,x_0}\equiv w^{(1)}_{0,k,-1,x_0}=\tilde{w}^{(1)}_{0,k,+1,x_0}=
\sqrt{\phi_t m^2+\frac{\phi_t}{\phi_s}k^2}\left[1+\left(\frac{g_t}{\phi_t}+
\frac{g_s}{\phi_s}\frac{k^2}{\phi_sm^2+k^2}\right)x_0\right]
\end{equation}
are independent of the magnetic field. As we have seen in \cite{DiPiazza2},
this fact gives the transverse ground states a particular relevance in the
case of strong magnetic fields because their energies are much smaller than
that of the excited Landau levels that depend on $\mathbf{B}$. In particular,
we have shown that in this case the probability that a pair is created with
both the electron and the positron in a transverse ground state is much larger
than the other probabilities \cite{DiPiazza2}. For this reason we will only
calculate this probability and then we need to compute only the one-particle
transverse ground states corrected up to first order. As in the previous
paragraph, we will present the calculations of the first-order corrections to the
electron transverse ground states and we will quote the analogous results for
the positron transverse ground states.

The first order corrections to a given zero-order electron transverse ground
state come from the coupling of this state to the following classes of states:
\begin{enumerate}
\item the zero-order electron transverse ground states with the same energy
(all but the state to be corrected);
\item the zero-order electron transverse ground states with different energy;
\item the zero-order electron states which are not transverse ground states;
\item the zero-order positron states.
\end{enumerate}
Suppose, now, that we want to calculate the first-order corrections to the
state labeled by $J_0\equiv\{0,k,-1,x_0\}=\{R_0,x_0\}$. The states in the
first class are labeled by $\{R_0,x'_0\}$ with $x'_0\neq x_0$ because they
have the same energy of the $J_0$-state. But, all the contributions vanish
because the perturbation (\ref{U_1p_ti}) can not couple two eigenstates of
$\mathcal{H}^{(0)}(\mathbf{r},-i\boldsymbol{\partial})$ with the
same $n_d$ and two different
$x_0$ and $x'_0$. Similarly, the states in the second class are
characterized by $J'_0\equiv\{0,k',-1,x'_0\}$ with $k'\neq k$ and, since
$[\mathcal{I}(\mathbf{r},-i\boldsymbol{\partial}),\mathcal{P}_{\parallel}]=0$
they do not give any contributions. Instead, the contributions from the states
of the remaining two classes are, in general, different from zero then we write
the state $U^{(1)}_{J_0}(\mathbf{r})$ as
\begin{equation}
U^{(1)}_{J_0}(\mathbf{r})=U^{(0)}_{J_0}(\mathbf{r})+\sideset{}{'}\sum_{J'}
\mathsf{P}^{(1)}_{J_0,J'}U^{(0)}_{J'}(\mathbf{r})+
\sum_{J'}\mathsf{Q}^{(1)}_{J_0,J'}V^{(0)}_{J'}(\mathbf{r})
\end{equation}
where the primed sum does not include the transverse ground states and where \cite{Landau3}
\begin{align}
\mathsf{P}^{(1)}_{J_0,J'} &=\frac{1}{\varepsilon^{(0)}_k-w^{(0)}_{R'}}
\int d\mathbf{r}\sqrt{\phi^3_s}U^{(0)\dag}_{J'}(\mathbf{r})
\mathcal{I}(\mathbf{r},-i\boldsymbol{\partial})U^{(0)}_{J_0}(\mathbf{r}),\\
\mathsf{Q}^{(1)}_{J_0,J'} &=\frac{1}{\varepsilon^{(0)}_k+\tilde{w}^{(0)}_{R'}}
\int d\mathbf{r}\sqrt{\phi^3_s}V^{(0)\dag}_{J'}(\mathbf{r})
\mathcal{I}(\mathbf{r},-i\boldsymbol{\partial})U^{(0)}_{J_0}(\mathbf{r}).
\end{align}
In these equations we introduced the zero-order energies of the transverse
ground states $\varepsilon^{(0)}_k$ defined as [see Eqs. (\ref{w_0}) and (\ref{w_t_0})]
\begin{equation}
\label{E_0}
\varepsilon^{(0)}_k\equiv w^{(0)}_{R_0}=\tilde{w}^{(0)}_{\tilde{R}_0}=
\sqrt{\phi_t m^2+\frac{\phi_t}{\phi_s}k^2}
\end{equation}
with $\tilde{R}_0=\{0,k,+1\}$. We start by calculating the coefficients
$\mathsf{P}^{(1)}_{J_0,J'}$. From the expression (\ref{U_1p_ti}) of the
interaction Hamiltonian and from Eqs. (\ref{U^0_f}) and (\ref{U^P0_f}) we have
\begin{equation}
\begin{split}
\mathsf{P}^{(1)}_{J_0,J'} &=\frac{1}{\varepsilon^{(0)}_k-w^{(0)}_{R'}}
\sqrt{\frac{w^{(0)}_{R'}+\sqrt{\phi_t}m}{2w^{(0)}_{R'}}}
\sqrt{\frac{\varepsilon^{(0)}_k+\sqrt{\phi_t}m}{2\varepsilon^{(0)}_k}}\times\\
&\times\bigg\{\left[\sqrt{\phi_t}(g_P-g_M)m-\frac{g_P}{2}\big(w^{(0)}_{R'}+
\varepsilon^{(0)}_k\big)+g_E\varepsilon^{(0)}_k\right]\int d\mathbf{r}
\Phi^{\prime \dag}_{n',x'_0}(\mathbf{r})x\Phi'_{0,x_0}(\mathbf{r})+\\
&\qquad +\left[-\sqrt{\phi_t}(g_P-g_M)m-\frac{g_P}{2}\big(w^{(0)}_{R'}+
\varepsilon^{(0)}_k\big)+g_E\varepsilon^{(0)}_k\right]\frac{\phi_t}{\phi_s}\times\\
&\qquad \times\frac{1}{\varepsilon^{(0)}_k+\sqrt{\phi_t}m}\frac{1}{w^{(0)}_{R'}+
\sqrt{\phi_t}m}\int d\mathbf{r} \Phi^{\prime \dag}_{n',x'_0}(\mathbf{r})
\mathcal{V}'(\mathbf{r},-i\boldsymbol{\partial})x
\mathcal{V}'(\mathbf{r},-i\boldsymbol{\partial})\Phi'_{0,x_0}(\mathbf{r})\bigg\}.
\end{split}
\end{equation}
By using the orthonormal properties of the functions
$\Phi'_{n,x_0}(\mathbf{r})$ [see Eq. (\ref{ort_Phi})] and Eq. (\ref{Vp_x_Vp}) we obtain
\begin{equation}
\label{P_1_f}
\begin{split}
\mathsf{P}^{(1)}_{J_0,J'} &=\frac{1}{\varepsilon^{(0)}_k-w^{(0)}_{R'}}
\sqrt{\frac{w^{(0)}_{R'}+\sqrt{\phi_t}m}{2w^{(0)}_{R'}}}
\sqrt{\frac{\varepsilon^{(0)}_k+\sqrt{\phi_t}m}{2\varepsilon^{(0)}_k}}\times\\
&\times\bigg\{\left[\sqrt{\phi_t}(g_P-g_M)m-\frac{g_P}{2}\big(w^{(0)}_{R'}+
\varepsilon^{(0)}_k\big)+g_E\varepsilon^{(0)}_k\right]\frac{1}{\sqrt{2eB}}\delta_{n',+1}
\delta_{\sigma', -1}+\\
&\qquad +\left[-\sqrt{\phi_t}(g_P-g_M)m-\frac{g_P}{2}\big(w^{(0)}_{R'}+
\varepsilon^{(0)}_k\big)+g_E\varepsilon^{(0)}_k\right]\frac{\phi_t}{\phi_s}\times\\
&\qquad \times\frac{1}{\varepsilon^{(0)}_k+\sqrt{\phi_t}m}\frac{1}{w^{(0)}_{R'}+
\sqrt{\phi_t}m}\left[ik\delta_{n',0}\delta_{\sigma', +1}+
\frac{k^2}{\sqrt{2eB}}\delta_{n',+1}\delta_{\sigma', -1}\right]\bigg\}
\delta_{k',k}\delta_{x'_0,x_0}=\\
&=\left(\mathsf{B}^{(1)}_{k,x_0}\delta_{n',+1}
\delta_{\sigma', -1}-i\mathsf{C}^{(1)}_{k,x_0}\delta_{n',0}\delta_{\sigma', +1}\right)
\delta_{k',k}\delta_{x'_0,x_0}
\end{split}
\end{equation}
where we defined the coefficients
\begin{align}
\label{B_c}
\begin{split}
\mathsf{B}^{(1)}_{k,x_0} &\equiv\frac{1}{\mathcal{E}^{(0)}_k-
\varepsilon^{(0)}_k}\sqrt{\frac{\mathcal{E}^{(0)}_k+\sqrt{\phi_t}m}
{2\mathcal{E}^{(0)}_k}}\sqrt{\frac{\varepsilon^{(0)}_k+\sqrt{\phi_t}m}
{2\varepsilon^{(0)}_k}}\times\\
&\times\left\{\sqrt{\phi_t}(g_M-g_P)\frac{m}{\sqrt{2eB}}
\left[1-\frac{\phi_t}{\phi_s}\frac{k^2}{\big(\mathcal{E}^{(0)}_k+
\sqrt{\phi_t}m\big)\big(\varepsilon^{(0)}_k+\sqrt{\phi_t}m\big)}\right]\right.+\\
&\qquad\left.+\frac{1}{\sqrt{2eB}}\left[\frac{g_P}{2}\big(\mathcal{E}^{(0)}_k+
\varepsilon^{(0)}_k\big)-g_E\varepsilon^{(0)}_k\right]\left[1+
\frac{\phi_t}{\phi_s}\frac{k^2}{\big(\mathcal{E}^{(0)}_k+
\sqrt{\phi_t}m\big)\big(\varepsilon^{(0)}_k+\sqrt{\phi_t}m\big)}\right]\right\},
\end{split}\\
\label{C_c}
\begin{split}
\mathsf{C}^{(1)}_{k,x_0} &\equiv\frac{1}{\mathcal{E}^{(0)}_k-
\varepsilon^{(0)}_k}\sqrt{\frac{\mathcal{E}^{(0)}_k+\sqrt{\phi_t}m}
{2\mathcal{E}^{(0)}_k}}\sqrt{\frac{\varepsilon^{(0)}_k+\sqrt{\phi_t}m}
{2\varepsilon^{(0)}_k}}\times\\
&\times\left[\sqrt{\phi_t}(g_M-g_P)m-\frac{g_P}{2}\big(\mathcal{E}^{(0)}_k+
\varepsilon^{(0)}_k\big)+g_E\varepsilon^{(0)}_k\right]\frac{\phi_t}{\phi_s}
\frac{k}{\big(\mathcal{E}^{(0)}_k+\sqrt{\phi_t}m\big)\big(\varepsilon^{(0)}_k+
\sqrt{\phi_t}m\big)}
\end{split}
\end{align}
with [see Eqs. (\ref{w_0}) and (\ref{w_t_0})]
\begin{equation}
\label{E_1}
\mathcal{E}^{(0)}_k\equiv w^{(0)}_{0,k,+1}=w^{(0)}_{1,k,-1}=
\tilde{w}^{(0)}_{0,k,-1}=\tilde{w}^{(0)}_{1,k,+1}=\sqrt{\phi_t m^2+
\frac{\phi_t}{\phi_s}\left(k^2+2eB\right)}
\end{equation}
the zero-order energy of the first excited Landau level. In the same
way we can write the coefficients $\mathsf{Q}^{(1)}_{J_0,J'}$ as
\begin{equation}
\label{Q_1_f}
\begin{split}
\mathsf{Q}^{(1)}_{J_0,J'} &=\frac{\sigma'}{\varepsilon^{(0)}_k+
\tilde{w}^{(0)}_{R'}}\sqrt{\frac{\tilde{w}^{(0)}_{R'}+\sqrt{\phi_t}m}
{2w^{(0)}_{R'}}}\sqrt{\frac{\varepsilon^{(0)}_k+\sqrt{\phi_t}m}
{2\varepsilon^{(0)}_k}}\times\\
&\times\bigg\{\left[\sqrt{\phi_t}(g_P-g_M)m-\frac{g_P}{2}
\big(\varepsilon^{(0)}_k-\tilde{w}^{(0)}_{R'}\big)+g_E\varepsilon^{(0)}_k\right]
(-1)\sqrt{\frac{\phi_t}{\phi_s}}\frac{1}{\tilde{w}^{(0)}_{R'}+\sqrt{\phi_t}m}\times\\
&\qquad \times\left[-i\delta_{n',0}\delta_{\sigma', -1}-kx_0\delta_{n',0}
\delta_{\sigma', +1}-\frac{k}{\sqrt{2eB}}\delta_{n',+1}\delta_{\sigma',+1}\right]+\\
&\qquad +\left[-\sqrt{\phi_t}(g_P-g_M)m-\frac{g_P}{2}\big(\varepsilon^{(0)}_k-
w^{(0)}_{R'}\big)+g_E\varepsilon^{(0)}_k\right]\sqrt{\frac{\phi_t}{\phi_s}}
\frac{1}{\varepsilon^{(0)}_k+\sqrt{\phi_t}m}\times\\
&\qquad \times\left[-kx_0\delta_{n',0}\delta_{\sigma', +1}-
\frac{k}{\sqrt{2eB}}\delta_{n',+1}\delta_{\sigma', +1}\right]\bigg\}
\delta_{k',-k}\delta_{x'_0,x_0}=\\
&=\left(-\mathsf{D}^{(1)}_{k,x_0}\delta_{n',0}
\delta_{\sigma', +1}+i\mathsf{E}^{(1)}_{k,x_0}\delta_{n',0}
\delta_{\sigma', -1}-\mathsf{F}^{(1)}_{k,x_0}\delta_{n',+1}
\delta_{\sigma',+1}\right)\delta_{k',-k}\delta_{x'_0,x_0}
\end{split}
\end{equation}
with
\begin{align}
\label{D_c}
\mathsf{D}^{(1)}_{k,x_0} &\equiv\frac{1}{2\left(\varepsilon^{(0)}_k\right)^2}
\sqrt{\phi_t}(g_M-g_P)m\sqrt{\frac{\phi_t}{\phi_s}}kx_0,\\
\label{E_c}
\begin{split}
\mathsf{E}^{(1)}_{k,x_0} &\equiv
\frac{1}{\mathcal{E}^{(0)}_k+\varepsilon^{(0)}_k}
\sqrt{\frac{\mathcal{E}^{(0)}_k+\sqrt{\phi_t}m}{2\mathcal{E}^{(0)}_k}}
\sqrt{\frac{\varepsilon^{(0)}_k+\sqrt{\phi_t}m}{2\varepsilon^{(0)}_k}}\times\\
&\times\left[\sqrt{\phi_t}(g_M-g_P)m-\frac{g_P}{2}
\big(\mathcal{E}^{(0)}_k-\varepsilon^{(0)}_k\big)-g_E\varepsilon^{(0)}_k\right]
\sqrt{\frac{\phi_t}{\phi_s}}\frac{1}{\mathcal{E}^{(0)}_k+\sqrt{\phi_t}m},
\end{split}\\
\label{F_c}
\begin{split}
\mathsf{F}^{(1)}_{k,x_0} &\equiv
\frac{1}{\mathcal{E}^{(0)}_k+\varepsilon^{(0)}_k}
\sqrt{\frac{\mathcal{E}^{(0)}_k+\sqrt{\phi_t}m}{2\mathcal{E}^{(0)}_k}}
\sqrt{\frac{\varepsilon^{(0)}_k+\sqrt{\phi_t}m}{2\varepsilon^{(0)}_k}}
\sqrt{\frac{\phi_t}{\phi_s}}\frac{k}{\sqrt{2eB}}\times\\
&\times\left\{\sqrt{\phi_t}(g_M-g_P)m\left(\frac{1}{\varepsilon^{(0)}_k+
\sqrt{\phi_t}m}+\frac{1}{\mathcal{E}^{(0)}_k+\sqrt{\phi_t}m}\right)+\right.\\
&\qquad\left.+\left[\frac{g_P}{2}\big(\mathcal{E}^{(0)}_k-
\varepsilon^{(0)}_k\big)+g_E\varepsilon^{(0)}_k\right]
\left(\frac{1}{\varepsilon^{(0)}_k+\sqrt{\phi_t}m}-
\frac{1}{\mathcal{E}^{(0)}_k+\sqrt{\phi_t}m}\right)\right\}.
\end{split}
\end{align}
If we also define the coefficients $\mathsf{A}^{(1)}_{k,x_0}$ as
\begin{equation}
\label{A_c}
\mathsf{A}^{(1)}_{k,x_0}\equiv\frac{g_E}{2}x_0
\end{equation}
then, the first order transverse ground state $U^{(1)}_{0,k,-1,x_0}
(\mathbf{r})$ can be written simply as
\begin{equation}
\label{U^1_f}
\begin{split}
U^{(1)}_{0,k,-1,x_0}(\mathbf{r}) &=\left(1+\mathsf{A}^{(1)}_{k,x_0}\right)
U^{(0)}_{0,k,-1,x_0}(\mathbf{r})+\mathsf{B}^{(1)}_{k,x_0}U^{(0)}_{1,k,-1,x_0}
(\mathbf{r})-i\mathsf{C}^{(1)}_{k,x_0}U^{(0)}_{0,k,+1,x_0}(\mathbf{r})-\\
&-\mathsf{D}^{(1)}_{k,x_0}V^{(0)}_{0,-k,+1,x_0}(\mathbf{r})+
i\mathsf{E}^{(1)}_{k,x_0}V^{(0)}_{0,-k,-1,x_0}(\mathbf{r})-
\mathsf{F}^{(1)}_{k,x_0}V^{(0)}_{1,-k,+1,x_0}(\mathbf{r}).
\end{split}
\end{equation}
The term $\mathsf{A}^{(1)}_{k,x_0}U^{(0)}_{0,k,-1,x_0}(\mathbf{r})=
g_Ex_0U^{(0)}_{0,k,-1,x_0}(\mathbf{r})/2$ has been added to compensate
for the factor $(1-g_Ex_0)$ in the scalar product (\ref{s_p_f}) and
then to have the states correctly normalized up to first order, as
\begin{equation}
(U^{(1)}_{J_0},U^{(1)}_{J'_0})=\delta_{J_0,J'_0}.
\end{equation}
We observe that, even if $\mathsf{A}^{(1)}_{k,x_0},\ldots,
\mathsf{F}^{(1)}_{k,x_0}$ are all first order-quantities in the
couplings $g_E$, $g_P$ and $g_M$, the coefficients $\mathsf{B}^{(1)}_{k,x_0}$,
$\mathsf{C}^{(1)}_{k,x_0}$, $\mathsf{E}^{(1)}_{k,x_0}$ and
$\mathsf{F}^{(1)}_{k,x_0}$ go, for strong magnetic fields, as
$1/\sqrt{2eB}$ and this circumstance makes them, in general, much smaller
than $\mathsf{A}^{(1)}_{k,x_0}$ and $\mathsf{D}^{(1)}_{k,x_0}$. We will
exploit this observation in the next paragraph.

Finally, with analogous calculations it can be shown that the first-order
positron transverse ground state $V^{(1)}_{0,k,+1,x_0}(\mathbf{r})$ can be written as
\begin{equation}
\label{V^1_f}
\begin{split}
V^{(1)}_{0,k,+1,x_0}(\mathbf{r}) &=\left(1+\mathsf{A}^{(1)}_{k,x_0}\right)
V^{(0)}_{0,k,+1,x_0}(\mathbf{r})+\mathsf{B}^{(1)}_{k,x_0}V^{(0)}_{1,k,+1,x_0}
(\mathbf{r})-i\mathsf{C}^{(1)}_{k,x_0}V^{(0)}_{0,k,-1,x_0}(\mathbf{r})-\\
&-\mathsf{D}^{(1)}_{k,x_0}U^{(0)}_{0,-k,-1,x_0}(\mathbf{r})+
i\mathsf{E}^{(1)}_{k,x_0}U^{(0)}_{0,-k,+1,x_0}(\mathbf{r})-
\mathsf{F}^{(1)}_{k,x_0}U^{(0)}_{1,-k,-1,x_0}(\mathbf{r}).
\end{split}
\end{equation}
The states $V^{(1)}_{0,k,+1,x_0}(\mathbf{r})$ are also normalized as
\begin{equation}
(V^{(1)}_{J_0},V^{(1)}_{J'_0})=\delta_{J_0,J'_0}
\end{equation}
and they are orthogonal up to first-order to the states $U^{(1)}_{J_0}(\mathbf{r})$:
\begin{equation}
(V^{(1)}_{J_0},U^{(1)}_{J'_0})=(U^{(1)}_{J_0},V^{(1)}_{J'_0})=0.
\end{equation}

Before continuing our computation of the instantaneous eigenstates of
the Hamiltonian $H^{(1)}(t)$ we want to show a nice property about the
mean value of the velocity operator $\boldsymbol{\alpha}$ in the first-order
eigenstates (\ref{U^1_f}) and (\ref{V^1_f}). In particular, we will calculate
the mean value of the $x$-component of the velocity given by $v_x\equiv\alpha_x$
and of the component of the velocity in the axis orthogonal to the $x$-axis
and to the direction of $\mathbf{B}$ given by $v_{\perp}\equiv\alpha_y
\cos\vartheta-\alpha_z\sin\vartheta$ [see Eq. (\ref{B_ti})] in the
eigenstates $U^{(1)}_{J_0}(\mathbf{r})=U^{(1)}_{0,k,-1,x_0}(\mathbf{r})$
and  $V^{(1)}_{\tilde{J}_0}(\mathbf{r})=V^{(1)}_{0,k,+1,x_0}(\mathbf{r})$
\begin{align}
\left\langle U^{(1)}_{J_0}\big\vert v_x\big\vert U^{(1)}_{J_0}\right\rangle &
=\int d\mathbf{r}\sqrt{\phi_s^3}(1-g_Ex)U^{(1)\dag}_{0,k,-1,x_0}(\mathbf{r})
\alpha_xU^{(1)}_{0,k,-1,x_0}(\mathbf{r}),\\
\left\langle U^{(1)}_{J_0}\big\vert v_{\perp}\big\vert U^{(1)}_{J_0}\right\rangle
&=\int d\mathbf{r}\sqrt{\phi_s^3}(1-g_Ex)U^{(1)\dag}_{0,k,-1,x_0}(\mathbf{r})
(\alpha_y\cos\vartheta-\alpha_z\sin\vartheta)U^{(1)}_{0,k,-1,x_0}(\mathbf{r})
\end{align}
and
\begin{align}
\left\langle V^{(1)}_{\tilde{J}_0}\big\vert v_x\big\vert
V^{(1)}_{\tilde{J}_0}\right\rangle &=\int d\mathbf{r}\sqrt{\phi_s^3}(1-g_Ex)
V^{(1)\dag}_{0,k,+1,x_0}(\mathbf{r})\alpha_xV^{(1)}_{0,k,+1,x_0}(\mathbf{r}),\\
\left\langle V^{(1)}_{\tilde{J}_0}\big\vert v_{\perp}\big\vert
V^{(1)}_{\tilde{J}_0}\right\rangle &=\int d\mathbf{r}\sqrt{\phi_s^3}(1-g_Ex)
V^{(1)\dag}_{0,k,+1,x_0}(\mathbf{r})(\alpha_y\cos\vartheta-\alpha_z
\sin\vartheta)V^{(1)}_{0,k,+1,x_0}(\mathbf{r}).
\end{align}
Now, we proceed by considering only the first two mean values because
that concerning the positrons can be calculated in an analogous way. By
using the transformation properties (\ref{alpha_x_p})-(\ref{alpha_z_p}) and
the definitions $\alpha_{\pm}=(\alpha_x\pm i\alpha_y)/2$ we have
\begin{align}
\label{v_x_U}
\begin{split}
\left\langle U^{(1)}_{J_0}\big\vert v_x\big\vert U^{(1)}_{J_0}\right\rangle &
=\int d\mathbf{r}\sqrt{\phi_s^3}(1-g_Ex)U^{\prime(1)\dag}_{0,k,-1,x_0}(\mathbf{r})
\alpha_xU^{\prime(1)}_{0,k,-1,x_0}(\mathbf{r})=\\
&=\left\langle U^{\prime(1)}_{J_0}\big\vert\alpha_-\big\vert U^{\prime(1)}_{J_0}
\right\rangle+\left\langle U^{\prime(1)}_{J_0}\big\vert\alpha_+
\big\vert U^{\prime(1)}_{J_0}\right\rangle,
\end{split}\\
\label{v_y_U}
\begin{split}
\left\langle U^{(1)}_{J_0}\big\vert v_{\perp}\big\vert U^{(1)}_{J_0}\right\rangle &=
\int d\mathbf{r}\sqrt{\phi_s^3}(1-g_Ex)U^{\prime(1)\dag}_{0,k,-1,x_0}(\mathbf{r})
\alpha_yU^{\prime(1)}_{0,k,-1,x_0}(\mathbf{r})=\\
&=i\left(\left\langle U^{\prime(1)}_{J_0}\big\vert\alpha_-\big\vert U^{\prime(1)}_{J_0}
\right\rangle-\left\langle U^{\prime(1)}_{J_0}\big\vert\alpha_+
\big\vert U^{\prime(1)}_{J_0}\right\rangle\right)
\end{split}
\end{align}
where, by keeping only the terms up to first order
\begin{align}
\label{U_p_m}
\begin{split}
\left\langle U^{\prime(1)}_{J_0}\big\vert\alpha_-\big\vert U^{\prime(1)}_{J_0}\right\rangle
&=\int d\mathbf{r}\sqrt{\phi_s^3}U^{\prime(0)\dag}_{0,k,-1,x_0}(\mathbf{r})\alpha_-
\left[\mathsf{B}^{(1)}_{k,x_0}U^{(0)}_{1,k,-1,x_0}(\mathbf{r})-
i\mathsf{C}^{(1)}_{k,x_0}U^{(0)}_{0,k,+1,x_0}(\mathbf{r})+\right.\\
&\qquad\qquad\qquad\qquad\qquad\qquad\left.+i\mathsf{E}^{(1)}_{k,x_0}
V^{(0)\dag}_{0,-k,-1,x_0}(\mathbf{r})-\mathsf{F}^{(1)}_{k,x_0}
V^{(0)}_{1,-k,+1,x_0}(\mathbf{r})\right],
\end{split}\\
\label{U_p_p}
\begin{split}
\left\langle U^{\prime(1)}_{J_0}\big\vert\alpha_+\big\vert
U^{\prime(1)}_{J_0}\right\rangle &=\int d\mathbf{r}\sqrt{\phi_s^3}
\left[\mathsf{B}^{(1)}_{k,x_0}U^{(0)\dag}_{1,k,-1,x_0}(\mathbf{r})+
i\mathsf{C}^{(1)}_{k,x_0}U^{(0)\dag}_{0,k,+1,x_0}(\mathbf{r})-
i\mathsf{E}^{(1)}_{k,x_0}V^{(0)\dag}_{0,-k,-1,x_0}(\mathbf{r})-\right.\\
&\qquad\qquad\qquad\left.-\mathsf{F}^{(1)}_{k,x_0}V^{(0)\dag}_{1,-k,+1,x_0}
(\mathbf{r})\right]\alpha_+U^{\prime(0)}_{0,k,-1,x_0}(\mathbf{r}).
\end{split}
\end{align}
Now, in appendix D we indicate the technique to calculate these matrix
elements and the results are
\begin{equation}
\label{M_E_U_V}
\left\langle U^{\prime(1)}_{J_0}\big\vert\alpha_-\big\vert U^{\prime(1)}_{J_0}\right\rangle-
\left\langle U^{\prime(1)}_{J_0}\big\vert\alpha_+\big\vert U^{\prime(1)}_{J_0}\right\rangle=
i\frac{\sqrt{\phi_sm^2+k^2}}{2eB}\left(\frac{g_t}{\phi_t}+\frac{g_s}{\phi_s}
\frac{k^2}{\phi_sm^2+k^2}\right)
\end{equation}
then, from Eqs. (\ref{v_x_U}) and (\ref{v_y_U}) we obtain
\begin{align}
\left\langle U^{(1)}_{J_0}\big\vert v_x\big\vert U^{(1)}_{J_0}\right\rangle &=0,\\
\left\langle U^{(1)}_{J_0}\big\vert v_{\perp}\big\vert U^{(1)}_{J_0}\right\rangle &=
-\frac{1}{eB}\sqrt{\frac{\phi_s}{\phi_t}}\sqrt{\phi_tm^2+\frac{\phi_t}{\phi_s}k^2}
\left(\frac{g_t}{\phi_t}+\frac{g_s}{\phi_s}\frac{k^2}{\phi_sm^2+k^2}\right)
\end{align}
and, analogously,
\begin{align}
\left\langle V^{(1)}_{\tilde{J}_0}\big\vert v_x\big\vert V^{(1)}_{\tilde{J}_0}\right\rangle
&=0,\\
\left\langle V^{(1)}_{\tilde{J}_0}\big\vert v_{\perp}\big\vert V^{(1)}_{\tilde{J}_0}
\right\rangle &=\frac{1}{eB}\sqrt{\frac{\phi_s}{\phi_t}}\sqrt{\phi_tm^2+\frac{\phi_t}
{\phi_s}k^2}\left(\frac{g_t}{\phi_t}+\frac{g_s}{\phi_s}\frac{k^2}{\phi_sm^2+k^2}\right).
\end{align}
This results are just the quantum counterpart of the so-called $(\mathbf{E}
\times\mathbf{B})$-drift-velocity effect typical of the motion of a charged
particle in the presence of a uniform and constant electromagnetic field
$(\mathbf{E},\mathbf{B})$ \cite{Landau2}. In our case, obviously, the electric
force is substituted by the gravitational force [see Eq. (\ref{F_g_x})] and the
sign of the drift velocity depends on the sign of the charge of the particle.
%
\subsection{Instantaneous first-order eigenstates and eigenvalues of the
Hamiltonian $H^{(1)}(t)$}
\label{III_3}

The results obtained in the previous paragraphs will be used here to solve our
initial problem: the determination of the instantaneous eigenstates and
eigenenergies of the slowly varying Hamiltonian
\begin{equation}
\label{H_tot_f_2}
\begin{split}
H^{(1)}(t) &=\int d\mathbf{r}\sqrt{\phi_s^3}(1-g_Ex)\Psi^{\dag}(\mathbf{r},t)
\mathcal{H}^{(1)}(\mathbf{r},-i\boldsymbol{\partial},t)\Psi(\mathbf{r},t).
\end{split}
\end{equation}
In fact, we have only to substitute everywhere in the results just obtained the
constant magnetic field $\mathbf{B}$ with the time-dependent magnetic field
$\mathbf{B}(t)$ given in Eq. (\ref{B}). In this way, the one-particle eigenstates
both at zero and first order will depend explicitly on time such as the
eigenenergies apart from those of the transverse ground states that do not
depend on the magnetic field.

With this prescription if we expand the field $\Psi (\mathbf{r},t)$ with
respect to the first-order instantaneous basis $[U^{(1)}_J(\mathbf{r},t),
V^{(1)}_J(\mathbf{r},t)]$ as
\begin{equation}
\Psi (\mathbf{r},t)=\sum_J \left[c^{(1)}_J(t) U^{(1)}_J(\mathbf{r},t)+
d_J^{(1)\dag}(t) V^{(1)}_J(\mathbf{r},t)\right]
\end{equation}
the instantaneous eigenstates of the Hamiltonian (\ref{H_tot_f}) are given
by the Fock states
\begin{equation}
|\{n_J\}(t);\{\tilde{n}_{\tilde{J}}\}(t)\rangle\equiv
\left[c^{(1)\dag}_{J_1}(t)\right]^{n_{\jmath_1}}
\left[c^{(1)\dag}_{J_2}(t)\right]^{n_{\jmath_2}}\cdots
\left[d^{(1)\dag}_{\tilde{J}_1}(t)\right]^{\tilde{n}_{\tilde{J}_1}}
\left[d^{(1)\dag}_{\tilde{J}_2}(t)\right]^{\tilde{n}_{\tilde{J}_2}}\cdots|0(t)\rangle
\end{equation}
with $|0(t)\rangle$ the instantaneous vacuum state, while the instantaneous eigenenergies are
\begin{equation}
E^{(1)}(t)=\sum_l \left[w^{(1)}_{J_l}(t)n_{J_l}+\tilde{w}^{(1)}_{\tilde{J}_l}(t)
\tilde{n}_{\tilde{J}_l}\right].
\end{equation}
In this framework the creation of a pair at time $t$ with the electron
in the state labeled by $J$ and the positron in the state labeled by
$\tilde{J}'$ is the transition from the vacuum state $|0(t)\rangle$ to
the pair state $|1_J(t);\tilde{1}_{\tilde{J}'}(t)\rangle\equiv c_J^{\dag}(t)
d_{\tilde{J}'}^{\dag}(t)|0(t)\rangle$.
%
%
\section{Calculation of the production probabilities}
\label{IV}
\setcounter{equation}{0}
\renewcommand{\theequation}{IV.\arabic{equation}}

As we have said in the previous section we can calculate the probability of
producing an $e^--e^+$ pair in the presence of the slowly varying magnetic
field (\ref{B}) and of the static gravitational field described by the metric
tensor (\ref{g_mu_nu_l}) by means of the adiabatic perturbation theory up to
first order in the time-derivative of the magnetic field. In order to avoid
the possible confusion between the ``first order'' relative to the adiabatic
perturbation theory and the ``first order'' relative to the gravitational
couplings $g_E$, $g_P$ and $g_M$, in what follows we will always refer to
the second one. In particular, the symbol $^{(1)}$ indicates quantities that
are first-order in the gravitational couplings.

From now on, the gravitational field will not play any further role: we took
into account its presence by correcting the one-particle electron and positron
eigenstates and eigenenergies. Now, following the adiabatic perturbation theory,
the operator responsible of the pair production is the time derivative of the
second-quantized Hamiltonian (\ref{H_tot_f}) that, by using Eq. (\ref{H_1p}),
can be written as
\begin{equation}
\label{H_dot_1}
\begin{split}
\dot{H}^{(1)}(t) &=\int d\mathbf{r}\sqrt{\phi_s^3}(1-g_Ex)\Psi^{\dag}(\mathbf{r},t)
\boldsymbol{\partial}_{\mathbf{B}}\mathcal{H}^{(1)}(\mathbf{r},
-i\boldsymbol{\partial},t)\cdot\dot{\mathbf{B}}(t)\Psi(\mathbf{r},t)=\\
&=\int d\mathbf{r}\sqrt{\phi_s^3}(1-g_Ex)\Psi^{\dag}(\mathbf{r},t)
\sqrt{\frac{\phi_t}{\phi_s}}\left[1-(g_P-g_E)x\right]\frac{e}{2}
\left(\mathbf{r}\times\boldsymbol{\alpha}\right)\cdot\dot{\mathbf{B}}(t)
\Psi(\mathbf{r},t)\simeq\\
&\simeq\sqrt{\frac{\phi_t}{\phi_s}}\frac{e\dot{\mathbf{B}}(t)}{2}\cdot\int
d\mathbf{r}\sqrt{\phi_s^3}(1-g_Px)\Psi^{\dag}(\mathbf{r},t)\left(\mathbf{r}
\times\boldsymbol{\alpha}\right)\Psi(\mathbf{r},t).
\end{split}
\end{equation}
Starting from the second line of this equation the physical
meaning of the result can be understood. In fact, the vector
$\boldsymbol{\alpha}$ can be interpreted as
the one-particle relativistic operator corresponding
to the velocity of the electron, then the quantity
$(\mathbf{r}\times\boldsymbol{\alpha})\cdot\dot{\mathbf{B}}(t)=-
[\mathbf{r}\times\dot{\mathbf{B}}(t)]\cdot\boldsymbol{\alpha}$ is
proportional to the scalar product of the external electric field
$-\partial\mathbf{A}(\mathbf{r},t)/\partial t$
[see Eq. (\ref{A})] and of the electron velocity that is to the work per unit
time done by the induced electric field itself. On this respect, we want to do a couple of observations that can also be referred to our previous papers \cite{Calucci,DiPiazza1,DiPiazza2}. Firstly, the induced electric field $\mathbf{E}(\mathbf{r},t)=-\partial\mathbf{A}(\mathbf{r},t)/\partial t$ is always perpendicular to the magnetic field $\mathbf{B}(t)$. Secondly, due to the presence of $\mathbf{E}(\mathbf{r},t)$ we may conclude that this last field is the responsible for the production according to the well-known mechanism
proposed by Schwinger in \cite{Schwinger}. There are however two differences to be stressed:
\begin{enumerate}
\item because of the presence of the time depending magnetic field $\mathbf{B}(t)$ the electric field $\mathbf{E}(\mathbf{r},t)$ is rotational, so it does not admit a scalar
potential and the interpretation of pair production as tunnel effect is
not straightforward;
\item the one-particle states of the produced electron-positron pair are states in a magnetic field and they very different from the one-particle states in an electric field (for example, the transverse motion here is quantized and the energy levels with different $n_d$ are well separated from each other).
\end{enumerate}

Now, we pointed out in the previous section that the probability that a pair is
created with both the electron and positron in a transverse ground state is
much larger than the others probabilities. If the electron is assumed to be
created in the transverse ground state state $U_{J_0}^{(1)}(\mathbf{r},t)$
with $J_0=\{0,k,-1,x_0\}$ and the positron in the transverse ground state
$V_{\tilde{J}'_0}^{(1)}(\mathbf{r},t)$ with $\tilde{J}'_0=\{0,k',+1,x'_0\}$,
then the matrix element of the transition is given by
\begin{equation}
\label{M_E_cr}
\begin{split}
\dot{H}^{(1)}_{J_0\tilde{J}'_0}(t)&\equiv\langle 1_{J_0}(t);\tilde{1}_{\tilde{J}'_0}(t)|
\dot{H}^{(1)}(t)|0(t)\rangle=\\
&=\sqrt{\frac{\phi_t}{\phi_s}}\frac{e\dot{\mathbf{B}}(t)}{2}\cdot\int d\mathbf{r}
\sqrt{\phi_s^3}(1-g_Px)U_{J_0}^{(1)\dag}(\mathbf{r},t)\left(\mathbf{r}\times
\boldsymbol{\alpha}\right)V^{(1)}_{\tilde{J}'_0}(\mathbf{r},t)
\end{split}
\end{equation}
where $|0(t)\rangle$ and $|1_{J_0}(t);\tilde{1}_{\tilde{J}'_0}(t)\rangle$ are the
vacuum and the pair state at time $t$ respectively. The creation amplitude at time
$t$ can be calculated from this matrix element as \cite{Migdal}
\begin{equation}
\label{cr_ampl}
\gamma^{(1)}_{J_0\tilde{J}'_0}(t)=\frac{1}{\varepsilon^{(1)}_{k,x_0}+
\varepsilon^{(1)}_{k',x'_0}}\int_0^tdt'\dot{H}^{(1)}_{J_0\tilde{J}'_0}(t')
\exp\left[i\left(\varepsilon^{(1)}_{k,x_0}+\varepsilon^{(1)}_{k',x'_0}\right)t'\right]
\end{equation}
where we used the fact that the first-order energies $\varepsilon^{(1)}_{k,x_0}$
of the transverse ground states do not depend on $\mathbf{B}(t)$ and then on
time [see Eq. (\ref{w_g_1})].

In the following we will distinguish two different time evolutions of the magnetic field:
\begin{enumerate}
\item rotating magnetic field:  $\mathbf{B}(t)=\mathbf{B}_{\curvearrowright}(t)=
B_{\curvearrowright}(0,\sin\omega t, \cos\omega t)$;
\item magnetic field varying only in strength: $\mathbf{B}(t)=\mathbf{B}_{\uparrow}(t)=
(0,0, B_{\uparrow}(t))$
\end{enumerate}
with
\begin{equation}
\label{B_2}
B_{\uparrow}(t)=B_f+(B_i-B_f)\exp\left(-\frac{t}{\tau}\right)
\end{equation}
and $B_i<B_f$. Obviously, in both cases the magnetic field is assumed to be strong
and slowly varying that is [see Eqs. (\ref{str_f}) and (\ref{sl_var})]
\begin{align}
B_{\curvearrowright} &\gg \frac{m^2}{e},\\
\omega &\ll m
\end{align}
and
\begin{align}
B_i &\gg \frac{m^2}{e},\\
\frac{B_f-B_i}{B_i}\frac{1}{\tau} &\ll m.
\end{align}
The first configuration describes better that case in which the magnetic field is produced by a rotating body while the second one concerns a situation in which it is generated by a collapsing body.
%
%
\subsection{Rotating magnetic field}

In the case of rotating magnetic fields we know that in Minkowski spacetime it is already
possible to create a pair with both the electron and the positron in a transverse ground
state \cite{DiPiazza2}. Our task here is to calculate the corrections due to the presence
of the gravitational field to this ``flat'' creation probability. For this reason and
reminding that for strong magnetic fields the coefficients $\mathsf{A}^{(1)}_{k,x_0}$
and $\mathsf{D}^{(1)}_{k,x_0}$ are much larger than $\mathsf{B}^{(1)}_{k,x_0}$,
$\mathsf{C}^{(1)}_{k,x_0}$,  $\mathsf{E}^{(1)}_{k,x_0}$ and $\mathsf{F}^{(1)}_{k,x_0}$,
we can keep for simplicity only the terms proportional to $\mathsf{A}^{(1)}_{k,x_0}$
and $\mathsf{D}^{(1)}_{k,x_0}$ (apart from the zero-order term) in the expressions of
the first-order transverse ground states.\footnote{Note that in the case of purely
rotating magnetic field the coefficients $\mathsf{A}^{(1)}_{k,x_0},\ldots,
\mathsf{F}^{(1)}_{k,x_0}$ are time-independent because they depend on the
strength of the magnetic field. The same thing can be said about the electron
and positron eigenenergies and about the rotated eigenstates
$U^{\prime(1)}_J(\mathbf{r})$ and $V^{\prime(1)}_J(\mathbf{r})$. Instead,
the eigenstates $U^{(1)}_J(\mathbf{r},t)$ and $V^{(1)}_J(\mathbf{r},t)$ depend
on time through the rotation operator $\mathcal{R}_x[\vartheta(t)]=
\mathcal{R}_x(\omega t)$.} With this simplification the matrix element (\ref{M_E_cr})
can be written as \cite{DiPiazza2}
\begin{equation}
\begin{split}
\dot{H}^{(1)}_{\curvearrowright;(J_0\tilde{J}'_0)}(t) &=\sqrt{\frac{\phi_t}{\phi_s}}
\frac{e\dot{\mathbf{B}}_{\curvearrowright}(t)}{2}\cdot\\
&\cdot\int d\mathbf{r}\sqrt{\phi_s^3}(1-g_Px)U_{J_0}^{\prime(1)\dag}(\mathbf{r})
\mathcal{R}_x(\omega t)\left(\mathbf{r}\times\boldsymbol{\alpha}\right)
\mathcal{R}^{\dag}_x(\omega t)V^{\prime(1)}_{\tilde{J}'_0}(\mathbf{r})=\\
&=
\sqrt{\frac{\phi_t}{\phi_s}}\frac{e\omega B_{\curvearrowright}}{2}\int
d\mathbf{r}\sqrt{\phi_s^3}(1-g_Px)U_{J_0}^{\prime(1)\dag}(\mathbf{r})
\left(\mathbf{r}\times\boldsymbol{\alpha}\right)_y
V^{\prime(1)}_{\tilde{J}'_0}(\mathbf{r})\simeq\\
&\simeq\sqrt{\frac{\phi_t}{\phi_s}}\frac{e\omega B_{\curvearrowright}}{2}
\int d\mathbf{r}\sqrt{\phi_s^3}(1-g_Px)\times\\
&\times\left[\left(1+\mathsf{A}^{(1)}_{k,x_0}\right)U^{\prime(0)\dag}_{0,k,-1,x_0}
(\mathbf{r})-\mathsf{D}^{(1)}_{k,x_0}V^{\prime(0)\dag}_{0,-k,+1,x_0}(\mathbf{r})\right]
x\alpha_z\times\\
&\times\left[\left(1+\mathsf{A}^{(1)}_{k',x'_0}\right)V^{\prime(0)}_{0,k',+1,x'_0}
(\mathbf{r})-\mathsf{D}^{(1)}_{k',x'_0}U^{\prime(0)}_{0,-k',-1,x'_0}(\mathbf{r})\right].
\end{split}
\end{equation}
The matrix element does not depend on time and, if we keep only the first order
terms in $g_E$, $g_P$ and $g_M$, we have
\begin{equation}
\label{M_E_cr_rot}
\begin{split}
\dot{H}^{(1)}_{\curvearrowright;(J_0\tilde{J}'_0)} &\simeq \sqrt{\frac{\phi_t}{\phi_s}}
\frac{e\omega B_{\curvearrowright}}{2}\bigg\{\int d\mathbf{r}\sqrt{\phi_s^3}U^{\prime(0)\dag}_{0,k,-1,x_0}
(\mathbf{r})(1-g_Px)x\alpha_zV^{\prime(0)}_{0,k',+1,x'_0}(\mathbf{r})+\\
&+\int d\mathbf{r}\sqrt{\phi_s^3}\left[\mathsf{A}^{(1)}_{k,x_0}U^{\prime(0)\dag}_{0,k,-1,x_0}
(\mathbf{r})-\mathsf{D}^{(1)}_{k,x_0}V^{\prime(0)\dag}_{0,-k,+1,x_0}(\mathbf{r})\right]
x\alpha_zV^{\prime(0)}_{0,k',+1,x'_0}(\mathbf{r})+\\
&+\int d\mathbf{r}\sqrt{\phi_s^3}U^{\prime(1)\dag}_{0,k,-1,x_0}(\mathbf{r})x\alpha_z
\left[\mathsf{A}^{(1)}_{k',x'_0}V^{\prime(0)}_{0,k',+1,x'_0}(\mathbf{r})-
\mathsf{D}^{(1)}_{k',x'_0}U^{\prime(0)}_{0,-k',-1,x'_0}(\mathbf{r})\right]\bigg\}.
\end{split}
\end{equation}
Now, from Eq. (\ref{x}) we see that the matrix elements of the operator $(1-g_Px)x$
between two transverse ground states that have both $n=0$ are equal to that of the
operator $(1-g_Px_0)x_0+1/(2eB)$ or, by coherently neglecting the last term, to that
of $(1-g_Px_0)x_0$. In this way, the matrix element (\ref{M_E_cr_rot}) becomes
\begin{equation}
\label{M_E_cr_rot_2}
\begin{split}
\dot{H}^{(1)}_{\curvearrowright;(J_0\tilde{J}'_0)} \simeq \sqrt{\frac{\phi_t}{\phi_s}}
\frac{e\omega B_{\curvearrowright}}{2}\bigg\{ &\left[1-(g_P-g_E)x_0\right]x_0\times\\
&\times\int d\mathbf{r}\sqrt{\phi_s^3}U^{\prime(0)\dag}_{0,k,-1,x_0}(\mathbf{r})
\alpha_zV^{\prime(0)}_{0,k',+1,x'_0}(\mathbf{r})-\\
&-\mathsf{D}^{(1)}_{k,x_0}x_0\int d\mathbf{r}\sqrt{\phi_s^3}
V^{\prime(0)\dag}_{0,-k,+1,x_0}(\mathbf{r})\alpha_zV^{\prime(0)}_{0,k',+1,x'_0}(\mathbf{r})-\\
&-\mathsf{D}^{(1)}_{k',x'_0}x_0\int d\mathbf{r}\sqrt{\phi_s^3}
U^{\prime(0)\dag}_{0,k,-1,x_0}(\mathbf{r})\alpha_z U^{\prime(0)}_{0,-k',-1,x'_0}
(\mathbf{r})\bigg\}.
\end{split}
\end{equation}
By using the explicit expressions of the transverse ground states given in
appendix C it can be easily seen that
\begin{align}
\int d\mathbf{r}\sqrt{\phi_s^3}U^{\prime(0)\dag}_{0,k,-1,x_0}(\mathbf{r})
\alpha_zV^{\prime(0)}_{0,k',+1,x'_0}(\mathbf{r}) &=-\frac{\sqrt{\phi_t}m}
{\varepsilon^{(0)}_k}\delta_{k,-k'}\delta_{x_0,x'_0},\\
\int d\mathbf{r}\sqrt{\phi_s^3}V^{\prime(0)\dag}_{0,-k,+1,x_0}(\mathbf{r})
\alpha_zV^{\prime(0)}_{0,k',+1,x'_0}(\mathbf{r})&=-\sqrt{\frac{\phi_t}{\phi_s}}
\frac{k}{\varepsilon^{(0)}_k}\delta_{k,-k'}\delta_{x_0,x'_0},\\
\int d\mathbf{r}\sqrt{\phi_s^3} U^{\prime(0)\dag}_{0,k,-1,x_0}(\mathbf{r})
\alpha_z U^{\prime(0)}_{0,-k',-1,x'_0}(\mathbf{r})&=\sqrt{\frac{\phi_t}{\phi_s}}
\frac{k}{\varepsilon^{(0)}_k}\delta_{k,-k'}\delta_{x_0,x'_0}.
\end{align}
By substituting these results and the expressions (\ref{D_c}) of the coefficients
$\mathsf{D}^{(1)}_{k,x_0}$ in Eq. (\ref{M_E_cr_rot_2}), we obtain
\begin{equation}
\label{M_E_cr_rot_f}
\dot{H}^{(1)}_{\curvearrowright;(J_0\tilde{J}'_0)}= -\sqrt{\frac{\phi_t}{\phi_s}}
\frac{e\omega B_{\curvearrowright}}{2}\frac{\sqrt{\phi_s}mx_0}
{\sqrt{\phi_s m^2+k^2}}\left[1+\left(\frac{g_t}{\phi_t}+
\frac{g_s}{\phi_s}\frac{\phi_s m^2}{\phi_s m^2+k^2}\right)x_0\right]
\delta_{k,-k'}\delta_{x_0,x'_0}
\end{equation}
where we used the definitions (\ref{g_E})-(\ref{g_M}) of $g_E$, $g_P$ and $g_M$.
Because of the presence of the Kronecker delta functions, the only transition
amplitude different from zero is given by [see Eq. (\ref{cr_ampl})]
\begin{equation}
\gamma^{(1)}_{\curvearrowright;(0,k,-1,x_0;0,-k,+1,x_0)}(t)=\frac{1}
{2\varepsilon^{(1)}_{k,x_0}}\dot{H}^{(1)}_{\curvearrowright;(J_0\tilde{J}'_0)}
\frac{\exp\left(i2\varepsilon^{(1)}_{k,x_0}t\right)-1}{2i\varepsilon^{(1)}_{k,x_0}}.
\end{equation}
If we square the modulus of this expression and multiply it by the number of
states [see Eq. (\ref{varrho})]
\begin{equation}
d\mathcal{N}_{\curvearrowright}=\frac{eB_{\curvearrowright}Y}{2\pi}dx_0\times\frac{Z}{2\pi}dk
\end{equation}
we obtain the differential probability that a pair is present at time $t$ with the electron
(positron) between $x_0$ and $x_0+dx_0$ and with longitudinal momentum between
$k$ and $k+dk$ ($-k$ and $-k-dk$):
\begin{equation}
\label{dP_rot}
\begin{split}
dP^{(1)}_{\curvearrowright}(x_0,k;t) &=\frac{eB_{\curvearrowright}}{16}
\frac{\phi_s}{\phi_t}\left(\frac{e\omega B_{\curvearrowright}}{2\pi}\right)^2
\frac{\phi_s m^2x_0^2}{(\phi_s m^2+k^2)^3}\left[1-\left(2\frac{g_t}{\phi_t}+\frac{g_s}{\phi_s}\frac{k^2-5\phi_s m^2}{\phi_s m^2+k^2}\right)x_0\right]\times\\
&\times \sin^2\left(\varepsilon^{(1)}_{k,x_0}t\right) dV^{(1)}dk
\end{split}
\end{equation}
where the continuum limits $Y\to\infty$ and $Z\to\infty$ have been tacitly
performed and where the ``physical'' first-order volume $dV^{(1)}=\phi_s^{3/2}(1-3g_sx_0/\phi_s)(YZdx_0)$ has been introduced. In conclusion, we obtain our final result by averaging Eq. (\ref{dP_rot})
with respect to time and by dividing it by the phase-space volume $dV^{(1)}dk$:
\begin{equation}
\label{dP_rot_f}
\begin{split}
\left\langle\frac{dP^{(1)}_{\curvearrowright}(x_0,k;t)}{dVdk}\right\rangle &=
\frac{1}{2\pi^2}\frac{\phi_s}{\phi_t}\left[\frac{e B_{\curvearrowright}}
{4\left(\phi_s m^2+k^2\right)}\right]^3\left(\sqrt{\phi_s} m\omega x_0\right)^2\times\\
&\times\left[1-\left(2\frac{g_t}{\phi_t}+\frac{g_s}{\phi_s}\frac{k^2-5\phi_s m^2}{\phi_s m^2+k^2}\right)x_0\right].
\end{split}
\end{equation}
This result is to be compared to the ``flat'' probability that can be obtained
by putting $g_t=g_s=0$ and $\phi_t=\phi_s=1$ [see Eqs. (\ref{Phi_c}) and (\ref{g})]
and that is identical to the result (1) in \cite{DiPiazza3}. We note that, since in our approximations $g_t/\phi_t$ can be assumed to be much larger than $g_s/\phi_s$, the creation probability increases for negative values of $x_0$. Analogously to the sign of the first-order
correction to the eigenenergies [see discussion below Eq. (\ref{w_1})], this fact
can be understood in classical terms by observing that the gravitational force
(\ref{F_g}) associated to the metric (\ref{g_mu_nu_l}) has only the negative
$x$-component (\ref{F_g_x}) and then, the electron and the positron are more
likely created at $x_0<0$. Nevertheless, we point out that the first-order
correction of the probability is not proportional to the gravitational force
but this is due to the fact that the operator responsible of the pair creation
is not the Hamiltonian but its time-derivative which has a completely different
structure. From this point of view, it is worth noting that
the physical origin of the presence of the length $x_0$ in Eq. (\ref{dP_rot_f}) is
twofold. In fact, since, as we have said below Eq. (\ref{H_dot_1}), the matrix element
of the time-derivative of the Hamiltonian is proportional to
the matrix element of the power supplied by the external electric field which
grows linearly with the transverse spatial coordinates, then
the probability (\ref{dP_rot_f}) results proportional to $x_0^2$. Instead,
the first order correction due to the gravitational field is
proportional to $x_0$ because the gravitational force is uniform
[see Eq. (\ref{f_1})]. In this sense, the dependence of our final results on
the quantity $x_0$ is unavoidable since it is a direct consequence of the
(approximated) model we used that is of the assumed uniformity both of the
magnetic and of the gravitational field at a microscopic scale.
A more realistic model should
take into account that these fields are not actually uniform in the whole
space and that they vanish far from their source but the calculations
would be too complicated. Concerning this point, we want to stress that from the derivation 
itself of Eq. (\ref{dP_rot_f}), a clear physical meaning can be assigned to the quantity 
$x_0$. In fact, it can be interpreted as a typical length scale in which the magnetic 
(gravitational) field produced by the astrophysical compact object can be 
assumed to be uniform and this fact gives the possibility to do also a quantitative 
prediction from Eq. (\ref{dP_rot_f}).

A final note concerns the presence in Eq. (\ref{dP_rot_f}) of an
overall factor $\phi_s/\phi_t>1$ that increases the creation probability with
respect to its ``flat'' value.
%
%
\subsection{Magnetic field varying only in strength}

We have seen in \cite{DiPiazza2} that, if the magnetic field varies only in
strength it is impossible in the Minkowski spacetime to create a pair with both
the electron and the positron in a transverse ground state. Instead, we want to
show here that this process is allowed in the spacetime metric (\ref{g_mu_nu_l})
because of the corrections induced by the gravitational field on the transverse
ground states of the electron and of the positron. In fact, if we choose the same
electron and positron states used in the previous paragraph, the transition
matrix element (\ref{M_E_cr}) in this case becomes
\begin{equation}
\label{M_E_cr_fd}
\dot{H}^{(1)}_{\uparrow;(J_0\tilde{J}'_0)}(t)=\sqrt{\frac{\phi_t}{\phi_s}}
\frac{e\dot{B}_{\uparrow}(t)}{2}\int d\mathbf{r}\sqrt{\phi_s^3}(1-g_Px)
U_{J_0}^{\prime(1)\dag}(\mathbf{r},t)\left(\mathbf{r}\times
\boldsymbol{\alpha}\right)_zV^{\prime(1)}_{\tilde{J}'_0}(\mathbf{r},t)
\end{equation}
where we used the fact that, since the magnetic field lies in the $z$-direction
for every $t$ then $\mathcal{R}_x[\vartheta(t)]\equiv I$.

Now, the selection rule (50) in \cite{DiPiazza2} concerning the spin of the
transverse ground states allows us to conclude that for the zero-order transverse
ground states the following equalities hold:
\begin{align}
\label{U_z_V}
\int d\mathbf{r}U_{J_0}^{\prime(0)\dag}(\mathbf{r},t)\left(\mathbf{r}\times
\boldsymbol{\alpha}\right)_zV^{\prime(0)}_{\tilde{J}'_0}(\mathbf{r},t) &=0,\\
\int d\mathbf{r}U_{J_0}^{\prime(0)\dag}(\mathbf{r},t)\left(\mathbf{r}\times
\boldsymbol{\alpha}\right)_zU^{\prime(0)}_{J'_0}(\mathbf{r},t) &=0,\\
\label{V_z_V}
\int d\mathbf{r}V_{\tilde{J}_0}^{\prime(0)\dag}(\mathbf{r},t)\left(\mathbf{r}
\times\boldsymbol{\alpha}\right)_zV^{\prime(0)}_{\tilde{J}'_0}(\mathbf{r},t) &=0
\end{align}
where $J'_0\equiv\{0,k',-1,x'_0\}$ and $\tilde{J}'_0\equiv\{0,k',+1,x'_0\}$.
By exploiting these equations and by keeping only the terms up to first order
we can write Eq. (\ref{M_E_cr_fd}) as
\begin{equation}
\label{M_E_cr_fd_p}
\begin{split}
\dot{H}^{(1)}_{\uparrow;(J_0\tilde{J}'_0)}(t) &\simeq
\sqrt{\frac{\phi_t}{\phi_s}}\frac{e\dot{B}_{\uparrow}(t)}{2}
\int d\mathbf{r}\sqrt{\phi_s^3}
U_{0,k,-1,x_0}^{\prime(0)\dag}(\mathbf{r},t)\left(\mathbf{r}\times
\boldsymbol{\alpha}\right)_z\times\\
&\qquad\qquad\qquad\quad\times\left[\mathsf{B}^{(1)}_{k',x'_0}(t)
V^{\prime(0)}_{1,k',+1,x'_0}(\mathbf{r},t)-i\mathsf{C}^{(1)}_{k',x'_0}(t)
V^{\prime(0)}_{0,k',-1,x'_0}(\mathbf{r},t)+\right.\\
&\qquad\qquad\qquad\qquad\;\left.+i\mathsf{E}^{(1)}_{k',x'_0}(t)
U^{\prime(0)\dag}_{0,-k',+1,x'_0}(\mathbf{r},t)-
\mathsf{F}^{(1)}_{k',x'_0}(t)U^{\prime(0)\dag}_{1,-k',-1,x'_0}(\mathbf{r},t)\right]+\\
&+\sqrt{\frac{\phi_t}{\phi_s}}
\frac{e\dot{B}_{\uparrow}(t)}{2}\int d\mathbf{r}\sqrt{\phi_s^3}
\left[\mathsf{B}^{(1)}_{k,x_0}(t)U^{\prime(0)\dag}_{1,k,-1,x_0}
(\mathbf{r},t)+i\mathsf{C}^{(1)}_{k,x_0}(t)U^{\prime(0)\dag}_{0,k,+1,x_0}
(\mathbf{r},t)\right.-\\
&\qquad\qquad\qquad\qquad\qquad\quad\left.-i\mathsf{E}^{(1)}_{k,x_0}(t)
V^{\prime(0)\dag}_{0,-k,-1,x_0}(\mathbf{r},t)-\mathsf{F}^{(1)}_{k,x_0}(t)
V^{\prime(0)\dag}_{1,-k,+1,x_0}(\mathbf{r},t)\right]\times\\
&\qquad\qquad\qquad\quad\times\left(\mathbf{r}\times\boldsymbol{\alpha}\right)_z
V^{\prime(0)}_{0,k',+1,x'_0}(\mathbf{r},t).
\end{split}
\end{equation}
This expression can be further simplified if we use the definitions (\ref{x})
and (\ref{y}) of the operators $x$ and $y$. In fact, by using the
substitutions (\ref{sub_1})-(\ref{sub_2}) the operator $\left(\mathbf{r}
\times\boldsymbol{\alpha}\right)_z$ can be written as
\begin{equation}
\begin{split}
\left(\mathbf{r}\times\boldsymbol{\alpha}\right)_z &=x\alpha_y-y\alpha_x=
\left[x_0+\frac{1}{\sqrt{2eB}}(a+a^{\dag})\right]i(\alpha_--\alpha_+)-\\
&\qquad\qquad\qquad\quad-\left[y_0+\frac{i}{\sqrt{2eB}}(a-a^{\dag})\right]
(\alpha_-+\alpha_+)=\\
&=i\alpha_-\left(x_0+iy_0+\sqrt{\frac{2}{eB}}a^{\dag}\right)-i\alpha_+
\left(x_0-iy_0+\sqrt{\frac{2}{eB}}a\right).
\end{split}
\end{equation}
Now, from the expressions (\ref{U_gr_0}) and (\ref{V_gr_0}) of the transverse
ground states we conclude that only the operator $i\alpha_-(x_0+iy_0)$ gives a
non-vanishing contribution in the first integral of Eq. (\ref{M_E_cr_fd_p}) and
that, analogously, only the operator $-i\alpha_+(x_0-iy_0)$ gives a non-vanishing
contribution in the second one, then the matrix element (\ref{M_E_cr_fd_p}) can
be written as
\begin{equation}
\begin{split}
\dot{H}^{(1)}_{\uparrow;(J_0\tilde{J}'_0)}(t) &=i\sqrt{\frac{\phi_t}{\phi_s}}
\frac{e\dot{B}_{\uparrow}(t)}{2}\left[(x_0+iy_0)\left\langle U^{\prime(1)}_{J_0}(t)
\big\vert\alpha_-\big\vert V^{\prime(1)}_{\tilde{J}'_0}(t)\right\rangle -\right.\\
&\left.\qquad\qquad\qquad\quad-(x_0-iy_0)\left\langle U^{\prime(1)}_{J_0}(t)
\big\vert\alpha_+\big\vert V^{\prime(1)}_{\tilde{J}'_0}(t)\right\rangle\right]=\\
&=i\sqrt{\frac{\phi_t}{\phi_s}}\frac{e\dot{B}_{\uparrow}(t)}{2}\left\{\left[x_0+
\frac{1}{eB_{\uparrow}(t)}\frac{\partial}{\partial x_0}\right]\left\langle
U^{\prime(1)}_{J_0}(t)\big\vert\alpha_-\big\vert V^{\prime(1)}_{\tilde{J}'_0}(t)
\right\rangle-\right.\\
&\left.\qquad\qquad\qquad\quad-\left[x_0-\frac{1}{eB_{\uparrow}(t)}
\frac{\partial}{\partial x_0}\right]\left\langle U^{\prime(1)}_{J_0}(t)
\big\vert\alpha_+\big\vert V^{\prime(1)}_{\tilde{J}'_0}(t)\right\rangle\right\}
\end{split}
\end{equation}
where
\begin{align}
\label{M_-}
\begin{split}
\left\langle U^{\prime(1)}_{J_0}(t)\big\vert\alpha_-\big\vert
V^{\prime(1)}_{\tilde{J}'_0}(t)\right\rangle &=\int d\mathbf{r}
\sqrt{\phi_s^3}U_{0,k,-1,x_0}^{\prime(0)\dag}(\mathbf{r},t)\alpha_-\times\\
&\qquad\quad\times\left[\mathsf{B}^{(1)}_{k',x'_0}(t)V^{\prime(0)}_{1,k',+1,x'_0}
(\mathbf{r},t)-i\mathsf{C}^{(1)}_{k',x'_0}(t)V^{\prime(0)}_{0,k',-1,x'_0}
(\mathbf{r},t)+\right.\\
&\qquad\quad\left.+i\mathsf{E}^{(1)}_{k',x'_0}(t)U^{\prime(0)\dag}_{0,-k',+1,x'_0}
(\mathbf{r},t)-\mathsf{F}^{(1)}_{k',x'_0}(t)U^{\prime(0)\dag}_{1,-k',-1,x'_0}
(\mathbf{r},t)\right],
\end{split}\\
\label{M_+}
\begin{split}
\left\langle U^{\prime(1)}_{J_0}(t)\big\vert\alpha_+\big\vert
V^{\prime(1)}_{\tilde{J}'_0}(t)\right\rangle &=\int d\mathbf{r}\sqrt{\phi_s^3}
\left[\mathsf{B}^{(1)}_{k,x_0}(t)U^{\prime(0)\dag}_{1,k,-1,x_0}(\mathbf{r},t)+
i\mathsf{C}^{(1)}_{k,x_0}(t)U^{\prime(0)\dag}_{0,k,+1,x_0}(\mathbf{r},t)-\right.\\
&\qquad\qquad\qquad\left. -i\mathsf{E}^{(1)}_{k,x_0}(t)V^{\prime(0)\dag}_{0,-k,-1,x_0}
(\mathbf{r},t)-\mathsf{F}^{(1)}_{k,x_0}(t)V^{\prime(0)\dag}_{1,-k,+1,x_0}
(\mathbf{r},t)\right]\times\\
&\qquad\quad\times\alpha_+V^{\prime(0)}_{0,k',+1,x'_0}(\mathbf{r},t)
\end{split}
\end{align}
and where we used Eq. (\ref{y_0}). The calculation of these matrix elements (\ref{M_-})

and (\ref{M_+}) is quite tedious and the technique to perform it is sketched in
appendix D. In particular, it can be shown that
\begin{equation}
\left\langle U^{\prime(1)}_{J_0}(t)\big\vert\alpha_-\big\vert
V^{\prime(1)}_{\tilde{J}'_0}(t)\right\rangle=-
\left\langle U^{\prime(1)}_{J_0}(t)\big\vert\alpha_+\big\vert
V^{\prime(1)}_{\tilde{J}'_0}(t)\right\rangle=-i\frac{1}{2eB_{\uparrow}(t)}
\sqrt{\phi_t}(g_M-g_P)mk\delta_{k,-k'}\delta_{x_0,x'_0}
\end{equation}
and, for this reason the terms with the derivative with respect to $x_0$
cancel each other and the final expression of $\dot{H}^{(1)}_{\uparrow;(J_0J'_0)}(t)$ is
\begin{equation}
\label{M_E_cr_fd_f}
\dot{H}^{(1)}_{\uparrow;(J_0J'_0)}(t)=\sqrt{\frac{\phi_t}{\phi_s}}
\frac{\dot{B}_{\uparrow}(t)}{B_{\uparrow}(t)}\frac{1}{2\varepsilon^{(0)}_k}
\sqrt{\phi_t}(g_M-g_P)mkx_0\delta_{k,-k'}\delta_{x_0,x'_0}.
\end{equation}
We want to point out that the disappearance of the electron charge $-e$ is only a
consequence of the strong field approximation $B_{\uparrow}(t)\gg B_{cr}$. In
particular, it would be incoherent to conclude that this matrix element does not
vanish in the limit $e\to 0$ because in this limit $B_{cr}=m^2/e\to \infty$ and
the strong field condition can not be satisfied. Nevertheless, we will see that
the creation probability which is physically meaningful will contain $e$.

The creation amplitudes at time $t$ can be calculated by means of Eq. (\ref{cr_ampl})
and the only one different from zero is equal to
\begin{equation}
\gamma^{(1)}_{\uparrow;(0,k,-1,x_0;0,-k,+1,x_0)}(t)=\sqrt{\frac{\phi_t}{\phi_s}}
\frac{k}{4\left(\varepsilon^{(0)}_k\right)^2}\sqrt{\phi_t}(g_M-g_P)mx_0\int_0^tdt'
\frac{\dot{B}_{\uparrow}(t')}{B_{\uparrow}(t')} \exp\left(2i\varepsilon^{(0)}_kt'\right)
\end{equation}
where only the first-order terms in the gravitational couplings have been kept.
Now, by substituting the expression (\ref{B_2}) of the magnetic field $B_{\uparrow}(t)$
and by calculating its time derivative
\begin{equation}
\dot{B}_{\uparrow}(t)=\frac{B_f-B_i}{\tau}\exp\left(-\frac{t}{\tau}\right)
\end{equation}
the creation amplitude can be written as
\begin{equation}
\gamma^{(1)}_{\uparrow;(0,k,-1,x_0;0,-k,+1,x_0)}(t)=\sqrt{\frac{\phi_t}{\phi_s}}
\frac{k}{4\left(\varepsilon^{(0)}_k\right)^2}\sqrt{\phi_t}(g_M-g_P)mx_0
\int_0^{t/\tau}ds'\frac{\exp\left[-\left(1-2i\varepsilon^{(0)}_k\tau\right)s'\right]}
{\exp(-s')-\frac{B_f}{B_f-B_i}}
\end{equation}
with $s'=t'/\tau$ a dimensionless variable. Now, in our model $\tau$ is a macroscopic
time connected to the typical evolution times of a neutron star or of a black hole,
then we can safely assume that $\varepsilon^{(0)}_k\tau\gg 1$. This allows us to
give an asymptotic estimate of the remaining integral in the limit $t\to\infty$.
The result is
\begin{equation}
\gamma^{(1)}_{\uparrow;(0,k,-1,x_0;0,-k,+1,x_0)}(t\to\infty)\sim
\sqrt{\frac{\phi_t}{\phi_s}}\frac{k}{4\left(\varepsilon^{(0)}_k\right)^2}
\sqrt{\phi_t}(g_M-g_P)mx_0\frac{1}{2i\varepsilon^{(0)}_k\tau}\frac{B_f-B_i}{B_i}.
\end{equation}
Finally, by squaring the modulus of this expression and by multiplying it by
the number of states at $t\to\infty$
\begin{equation}
d\mathcal{N}_{\uparrow}(t\to\infty)=\frac{eB_{\uparrow}(t\to\infty)Y}{2\pi}dx_0
\times\frac{Z}{2\pi}dk=\frac{eB_f}{(2\pi)^2}dV^{(0)}dk
\end{equation}
with $dV^{(0)}=\sqrt{\phi_s^3}YZdx_0$ the ``physical'' quantization volume up to zero order [see Eq. (\ref{g_mu_nu_l_2})], we obtain the differential probability that a pair is created with the electron
(positron) between $x_0$ and $x_0+dx_0$ and with longitudinal momentum between $k$
and $k+dk$ ($-k$ and $-k-dk$)
\begin{equation}
\label{dP_fd}
dP^{(1)}_{\uparrow}(x_0,k;t\to\infty)\sim\frac{\phi_s}{\phi_t}\frac{\phi_s m^2k^2}
{(\phi_s m^2+k^2)^3}eB_f\left(\frac{B_f-B_i}{\tau B_i}\right)^2\left(\frac{g_sx_0}
{16\pi\phi_s}\right)^2dV^{(0)}dk.
\end{equation}
In this equation the continuum limits $Y\to\infty$ and $Z\to\infty$ have been performed and, since the probability is already proportional to $g_s$ and our calculations are exact up to first order in $g_s$ and $g_t$, it is enough to use the zero-order ``physical'' volume $dV^{(0)}$. Finally, the corresponding probability per unit volume and unit longitudinal momentum is given by
\begin{equation}
\label{dP_fd_f}
\frac{dP^{(1)}_{\uparrow}(x_0,k;t\to\infty)}{dV^{(0)}dk}\sim\left(\frac{g_s}{16\pi\phi_s}\right)^2
\frac{\phi_s}{\phi_t}\frac{eB_f}{(\phi_s m^2+k^2)^3}\left(\frac{B_f-B_i}{\tau B_i}\right)^2
\left(\sqrt{\phi_s} mx_0k\right)^2.
\end{equation}

As in the case of a rotating magnetic field we observe the presence of the overall
factors $\phi_s/\phi_t$ and $x_0^2$. Also, in \cite{Calucci} it
was calculated the total probability
that a pair is created in a slowly varying magnetic field with fixed direction but
in the Minkowski spacetime [see Eq. (20) in this paper]. We remind that in this
case either the electron or the positron must be created in a state which is not
a transverse ground state. In order to have a quantity to be compared to the
total probability given in Eq. (20) in \cite{Calucci}, we have to integrate Eq.
(\ref{dP_fd}) with respect to $x_0$ and to $k$. After these integrations it can
easily be seen that the order of magnitude of the ratio between our probability
and that given in \cite{Calucci} is\footnote{We have to be satisfied of an
order-of-magnitude comparison because the time evolution of the magnetic
field used in \cite{Calucci} is different from that used here.}
\begin{equation}
\eta\sim\frac{\phi_s}{\phi_t}\frac{1}{\sqrt{\phi_s}m}\left(\frac{g_s}{\phi_s}\right)^2
\frac{1}{\sqrt{eB_f}}.
\end{equation}
If we substitute the expressions (\ref{Phi_c}) and (\ref{g}) of $\phi_t$,
$\phi_s$ and $g_s$ we see that because of the inequality (\ref{ineq}) and of
the strong-field condition $eB_f\gg m^2$ then $\eta$ results much less than
one and then the gravitational effect is in our approximations small.
Nevertheless, the effect is there and it is reasonable to imagine that it can
be amplified in the presence of a real gravitational field which is not
restricted by our assumptions. For this reason we decided to consider also a particular case slightly different from the one at hand and where 
the gravitational field can be treated without approximations and this is the subject of the next paper. However, it must be pointed out that the perturbative approach gives the possibility to treat more general magnetic and gravitational field configurations while the strong field case the two fields must be necessarily parallel.
%
%
\section{Conclusions}
\label{V}

In this paper we have modified our previous results given in \cite{Calucci,DiPiazza1,DiPiazza2}
about the production of $e^--e^+$ pairs in the presence of strong, uniform and slowly
varying magnetic fields taking into account the presence of the gravitational field
represented by the metric tensor (\ref{g_mu_nu_l}). We point out that this metric tensor resulted from some approximations that we have done in section (\ref{II}) and that we resume for the sake of clarity: we have neglected the gravitational field due to the magnetic field energy, we have neglected the gravitational effects of the angular momentum of the stellar object and we have assumed to be not to close to the event horizon of the body.

We have treated the gravitational field perturbatively up to first-order in the
couplings $g_E$, $g_P$ and $g_M$ and we have seen how its presence modifies the
one-particle eigenstates and eigenenergies of the electron and of the positron.
This circumstance, obviously, changes the pair production probabilities we
calculated in absence of the gravitational field, \textit{i.e.} in the Minkowski
spacetime.

In particular, we have reexamined the case of a purely rotating magnetic field
\cite{DiPiazza2} and of a magnetic field varying only in strength \cite{Calucci}.
Firstly, in both cases we have found that the production probabilities contain an
amplification factor $\phi_s/\phi_t>1$ [see Eqs. (\ref{dP_rot_f}) and (\ref{dP_fd_f})].
Also, in the case of the purely rotating magnetic field we have seen that the
presence of the gravitational field in the plane orthogonal to the magnetic
field brakes the rotational symmetry in this plane and makes more likely that
a pair is created at $x_0<0$ [see Eq. (\ref{dP_rot_f})]. Instead, in the case
of a magnetic field with fixed direction we obtained a qualitative new result:
in the presence of the gravitational field it is possible to create a pair with
both the electron and the positron in a transverse ground state. Nevertheless,
we have seen at the end of the previous section that this probability is a small
quantity with respect to the total probability that a pair is created in the
Minkowski spacetime. But, this result is a consequence of the fact that the
gravitational field has been treated perturbatively and this treatment could
not fit what it happens, for example, near the event horizon of a black hole.
From this point of view, the information we have gained is that in the presence
of a gravitational field this new effect is there. In the next paper we analyze
a particular case in which the gravitational field can be treated without
approximations and we will show that in that case the effect is relevant.
%
%
\section*{Acknowledgments}
The author wishes to thank Prof. G. Calucci, Dr. E. Spallucci and Dr. S. Ansoldi for
stimulating discussions and useful advice. This work has been partially supported by
the Italian Ministry:
Ministero dell'Istruzione, Universit\`a e Ricerca, by means of the
Fondi per la Ricerca scientifica - Universit\`a di Trieste.

%
%
\appendix
\section*{Appendix A}
\setcounter{equation}{0}
\renewcommand{\theequation}{A\arabic{equation}}

In this appendix we want to quote some properties of the eigenstates $u_j(\mathbf{r})$
and $v_j(\mathbf{r})$ of the zero-order Hamiltonian (\ref{H_1p_0_ti}). Firstly, we
remind that $j\equiv\{n_d,k,\sigma,n_g\}$ embodies all the quantum numbers and that
$u_j(\mathbf{r})$ and $v_j(\mathbf{r})$ can be written in the form \cite{DiPiazza2}
\begin{align}
\label{xi_0}
u_j(\mathbf{r}) &=\mathcal{R}^{\dag}_x(\vartheta)u'_j(\mathbf{r}),\\
\label{eta_0}
v_j(\mathbf{r}) &=\mathcal{R}^{\dag}_x(\vartheta)v'_j(\mathbf{r})
\end{align}
where
\begin{equation}
\label{op_rot}
\mathcal{R}_x(\vartheta)=\exp(-i\vartheta \mathcal{J}_x)
\end{equation}
with $\mathcal{J}_x=\mathcal{L}_x+\mathcal{S}_x$ the $x$-component
of the one-particle total angular momentum operator and $\vartheta$ is defined
in Eq. (\ref{theta_ti}). The rotated spinors $u'_j(\mathbf{r})$ and $v'_j(\mathbf{r})$
are the solutions of Eqs. (\ref{Eq_eig_u_0}) and (\ref{Eq_eig_v_0}) with the vector
potential
\begin{equation}
\mathbf{A}'(\mathbf{r})=-\frac{1}{2}(\mathbf{r}\times\mathbf{B}')
\end{equation}
and
\begin{equation}
\mathbf{B}'=\left(
\begin{array}{c}
0\\
0\\
B
\end{array}\right)=\left(
\begin{array}{c}
0\\
0\\
\sqrt{B_y^2+B_z^2}
\end{array}\right)
\end{equation}
[see Eqs. (\ref{B_ti})-(\ref{A_ti})], and are given by \cite{DiPiazza2}
\begin{align}
\label{xi_pr_0}
u'_j(\mathbf{r}) &=\frac{1}{\sqrt[4]{\phi_s^3}}\sqrt{\frac{w^{(0)}_r+
\sqrt{\phi_t}m}{2w^{(0)}_r}}
\left(\begin{array}{c}
      \varphi'_j(\mathbf{r})\\ \sqrt{\dfrac{\phi_t}{\phi_s}}
\dfrac{\mathcal{V}'(\mathbf{r},-i\boldsymbol{\partial})}{w_r^{(0)}+
\sqrt{\phi_t}m}\varphi'_j(\mathbf{r})
      \end{array}\right), \\
\label{eta_pr_0}
v'_j(\mathbf{r}) &=\frac{\sigma}{\sqrt[4]{\phi_s^3}}\sqrt{\frac{\tilde{w}_q^{(0)}+
\sqrt{\phi_t}m}{2\tilde{w}_q^{(0)}}}
\left(\begin{array}{c}
      -\sqrt{\dfrac{\phi_t}{\phi_s}}\dfrac{\mathcal{V}'
(\mathbf{r},-i\boldsymbol{\partial})}{\tilde{w}_q^{(0)}+\sqrt{\phi_t}m}\chi'_j(\mathbf{r})\\
      \chi'_j(\mathbf{r})
      \end{array}\right)
\end{align}
where the indices $r$ and $q$ and the energies $w^{(0)}_r$ and $\tilde{w}_q^{(0)}$
have been defined in Eqs. (\ref{r}) and (\ref{q}) and in Eqs. (\ref{w_0}) and
(\ref{w_t_0}) respectively and where the numerical factor $1/\sqrt[4]{\phi_s^3}$
has been inserted to make the spinors correctly normalized with respect to the
scalar product (\ref{s_p_f}) at zero order:
\begin{align}
&\int d\mathbf{r}\sqrt{\phi_s^3}u^{\dag}_j(\mathbf{r})u_{j'}(\mathbf{r})=
\int d\mathbf{r}\sqrt{\phi_s^3}u^{\prime\dag}_j(\mathbf{r})u^{\prime}_{j'}(\mathbf{r})=
\delta_{j,j'},\\
&\int d\mathbf{r}\sqrt{\phi_s^3}v^{\dag}_j(\mathbf{r})v_j(\mathbf{r})=
\int d\mathbf{r}\sqrt{\phi_s^3}v^{\prime\dag}_j(\mathbf{r})v'_j(\mathbf{r})=\delta_{j,j'}
\end{align}
with $\delta_{j,j'}\equiv\delta_{n_d,n'_d}\delta_{k,k'}\delta_{\sigma,\sigma'}
\delta_{n_g,n'_g}$. Now, we give a more explicit expression of the operator
$\mathcal{V}'(\mathbf{r},-i\boldsymbol{\partial})$ and of the two-dimensional
spinors $\varphi'_j(\mathbf{r})$ and $\chi'_j(\mathbf{r})$ appearing in
Eqs. (\ref{xi_pr_0}) and (\ref{eta_pr_0}). To do this, we need to express the
operators $x$, $-i\partial/\partial x$, $y$ and $-i\partial/\partial y$ with
respect to the destruction and creation operators (that we indicate as $a_d$,
$a_d^{\dag}$, $a_g$ and $a_g^{\dag}$ respectively) corresponding to the quantum
numbers $n_d$ and $n_g$ and vice versa. These expressions can be found in
\cite{Cohen} and are given by
\begin{align}
\label{a_d}
a_d &=\frac{1}{2}\left[\sqrt{\frac{eB}{2}}(x-iy)+\sqrt{\frac{2}{eB}}
\left(\frac{\partial}{\partial x}+\frac{1}{i}\frac{\partial}{\partial y}\right)\right],\\
\label{a_d_c}
a_d^{\dag} &=\frac{1}{2}\left[\sqrt{\frac{eB}{2}}(x+iy)-\sqrt{\frac{2}{eB}}
\left(\frac{\partial}{\partial x}-\frac{1}{i}\frac{\partial}{\partial y}\right)\right],\\
\label{a_g}
a_g &=\frac{1}{2}\left[\sqrt{\frac{eB}{2}}(x+iy)+\sqrt{\frac{2}{eB}}
\left(\frac{\partial}{\partial x}-\frac{1}{i}\frac{\partial}{\partial y}\right)\right],\\
\label{a_g_c}
a_g^{\dag} &=\frac{1}{2}\left[\sqrt{\frac{eB}{2}}(x-iy)-\sqrt{\frac{2}{eB}}
\left(\frac{\partial}{\partial x}+\frac{1}{i}\frac{\partial}{\partial y}\right)\right]
\end{align}
and by
\begin{align}
\label{x}
x &=\frac{1}{2}\sqrt{\frac{2}{eB}}(a_g+a_g^{\dag}+a_d+a_d^{\dag})=x_0+
\frac{1}{\sqrt{2eB}}(a_d+a_d^{\dag}),\\
\frac{1}{i}\frac{\partial}{\partial x} &=\frac{1}{2i}\sqrt{\frac{eB}{2}}
(a_g-a_g^{\dag}+a_d-a_d^{\dag})=\frac{eB}{2}y_0+\frac{1}{2i}
\sqrt{\frac{eB}{2}}(a_d-a_d^{\dag}),\\
\label{y}
y &=\frac{1}{2i}\sqrt{\frac{2}{eB}}(a_g-a_g^{\dag}-a_d+a_d^{\dag})=y_0-
\frac{1}{i\sqrt{2eB}}(a_d-a_d^{\dag}),\\
\label{p_y}
\frac{1}{i}\frac{\partial}{\partial y} &=-\frac{1}{2}\sqrt{\frac{eB}{2}}
(a_g+a_g^{\dag}-a_d-a_d^{\dag})=-\frac{eB}{2}x_0+\frac{1}{2}\sqrt{\frac{eB}{2}}(a_d+a_d^{\dag})
\end{align}
where we introduced the operators
\begin{align}
\label{xy_perp}
x_0 &=\frac{1}{2}\sqrt{\frac{2}{eB}}(a_g+a_g^{\dag}),\\
y_0 &=\frac{1}{2i}\sqrt{\frac{2}{eB}}(a_g-a_g^{\dag}).
\end{align}
Starting from the well-known commutation relations among the operators $x$,
$-i\partial/\partial x$, $y$ and $-i\partial/\partial y$ it is easy to see that
\begin{align}
\label{comm_a}
[a_g,a_g^{\dag}] &=[a_d,a_d^{\dag}]=1,\\
[a_g,a_d] &=[a_g,a_d^{\dag}]=0
\end{align}
and that
\begin{equation}
\label{comm_x_0}
[x_0,y_0]=\frac{i}{eB}.
\end{equation}
In particular, this last commutator allows us to put
\begin{equation}
\label{y_0}
y_0=\frac{1}{eB}\frac{1}{i}\frac{\partial}{\partial x_0}.
\end{equation}

Now, the operator $\mathcal{V}'(\mathbf{r},-i\boldsymbol{\partial})$ in Eqs.
(\ref{xi_pr_0}) and (\ref{eta_pr_0}) is defined as [see Eq. (21) in \cite{DiPiazza2}]
\begin{equation}
\label{V_p_d}
\mathcal{V}'(\mathbf{r},-i\boldsymbol{\partial})\equiv\boldsymbol{\sigma}\cdot
\left[-i\boldsymbol{\partial}+e\mathbf{A}'(\mathbf{r})\right]
\end{equation}
then by substituting the expressions (\ref{x})-(\ref{p_y}) it can easily be shown that
\begin{equation}
\label{V_p}
\mathcal{V}'(\mathbf{r},-i\boldsymbol{\partial})=i\sqrt{2eB}(a_d^{\dag}\sigma_--
a_d\sigma_+)+\sigma_z\frac{1}{i}\frac{\partial}{\partial z}
\end{equation}
where $\sigma_{\pm}=(\sigma_x\pm i\sigma_y)/2$. It is also convenient here to
express the two-dimensional spinors $\varphi'_j(\mathbf{r})$ and $\chi'_j(\mathbf{r})$
in Cartesian coordinates instead of in cylindrical coordinates as in Eq. (22) of
Ref. \cite{DiPiazza2}:
\begin{align}
\label{phi}
\varphi'_j(\mathbf{r})
&=\frac{\exp(ikz)}{\sqrt{Z}}f'_{\sigma}\theta'_{n_d,n_g}(x,y),\\
\label{chi}
\chi'_j(\mathbf{r})
&=\frac{\exp(-ikz)}{\sqrt{Z}}f'_{-\sigma}\theta'_{n_g,n_d}(x,y).
\end{align}
In these expressions $Z$ is the length of the quantization volume in the $z$-direction,
\begin{align}
f'_{+1}=\left(\begin{array}{c}
             1\\
             0
             \end{array}\right), &&
f'_{-1}=\left(\begin{array}{c}
             0\\
             1
             \end{array}\right)
\end{align}
and where the scalar functions
\begin{equation}
\label{phi_t}
\theta'_{l_1,l_2}(x,y)=\sqrt{\frac{eB}{2\pi}\frac{1}{l_1!}\frac{1}{l_2!}}
(a_d^{\dag})^{l_1}(a_g^{\dag})^{l_2}\exp\left[-\frac{eB(x^2+y^2)}{4}\right]
\end{equation}
depend only on the transverse coordinates. In Eqs. (\ref{phi}) and (\ref{chi})
the operators $a_d^{\dag}$ and $a_g^{\dag}$ are supposed to be expressed as in
Eqs. (\ref{a_d_c}) and (\ref{a_g_c}) and from those equations it can be shown that
\begin{align}
(a_g^{\dag})^{n_g}\exp\left[-\frac{eB(x^2+y^2)}{4}\right] &=
\sqrt{\left(\frac{eB}{2}\right)^{n_g}}(x-iy)^{n_g}\exp\left[-\frac{eB(x^2+y^2)}{4}\right],\\
(a_d^{\dag})^{n_d}\exp\left[-\frac{eB(x^2+y^2)}{4}\right] &=
\sqrt{\left(\frac{eB}{2}\right)^{n_d}}(x+iy)^{n_d}\exp\left[-\frac{eB(x^2+y^2)}{4}\right]
\end{align}
and then that the transverse functions (\ref{phi_t}) can be written as
\begin{equation}
\label{phi_t_f}
\theta'_{l_1,l_2}(x,y)=\frac{1}{\sqrt{l_1!}}(a_d^{\dag})^{l_1}
\sqrt{\left(\frac{eB}{2}\right)^{l_2+1}\frac{1}{\pi l_2!}}(x-iy)^{l_2}
\exp\left[-\frac{eB(x^2+y^2)}{4}\right].
\end{equation}

Finally, we want to derive some orthonormalization properties of the
two-dimensional spinors $\varphi'_j(\mathbf{r})$ and $\chi'_j(\mathbf{r})$
and of the spinors $u_j(\mathbf{r})$ and $v_j(\mathbf{r})$. From the
commutation relations (\ref{comm_a}) it can easily be shown that, given
two quantum numbers $n_d$ and $n'_d$ with $n'_d\ge n_d$, then
\begin{equation}
(a_d)^{n_d}(a_d^{\dag})^{n'_d}=(a_da_d^{\dag})^{n_d}(a_d^{\dag})^{n'_d-n_d}=
(1+a_d^{\dag}a_d)^{n_d}(a_d^{\dag})^{n'_d-n_d}=\left[\sum_{l=0}^{n_d}
\binom{n_d}{l}(a_d^{\dag}a_d)^l\right](a_d^{\dag})^{n'_d-n_d}
\end{equation}
and, symmetrically,
\begin{equation}
(a_g)^{n_g}(a_g^{\dag})^{n'_g}=(a_ga_g^{\dag})^{n_g}(a_g^{\dag})^{n'_g-n_g}=
(1+a_g^{\dag}a_g)^{n_g}(a_g^{\dag})^{n'_g-n_g}=\left[\sum_{l=0}^{n_g}
\binom{n_g}{l}(a_g^{\dag}a_g)^l\right](a_g^{\dag})^{n'_g-n_g}
\end{equation}
if $n'_g\ge n_g$ (if $n'_d\le n_d$ or $n'_g\le n_g$ analogous relations are obtained).
With the help of these two equations and by reminding that $[a_g,a_d]=0$ it can be
seen that the following orthonormalization relations hold:
\begin{align}
\label{ort_phi_1}
\int d\mathbf{r}\varphi_j^{\prime\dag}(\mathbf{r})\varphi'_{j'}(\mathbf{r}) &=\delta_{j,j'},\\
\label{ort_chi_1}
\int d\mathbf{r}\chi_j^{\prime\dag}(\mathbf{r})\chi'_{j'}(\mathbf{r}) &=\delta_{j,j'}.
\end{align}
Now, if we calculate the square of the operator $\mathcal{V}'
(\mathbf{r},-i\boldsymbol{\partial})$ as given in Eq. (\ref{V_p}) we have
\begin{equation}
\label{V_p_2}
\begin{split}
\mathcal{V}^{\prime\, 2}(\mathbf{r},-i\boldsymbol{\partial}) &=-
\frac{\partial^2}{\partial z^2}+2eB\left[a_d^{\dag}a_d(\sigma_-\sigma_++
\sigma_+\sigma_-)+\sigma_+\sigma_-\right]=\\
&=-\frac{\partial^2}{\partial z^2}+eB\left(2a_d^{\dag}a_d+1+\sigma_z\right),
\end{split}
\end{equation}
then, from the expressions  (\ref{w_0}) and (\ref{w_t_0}) of the
zero-order eigenenergies, we obtain
\begin{align}
\label{ort_phi_2}
\frac{\phi_t}{\phi_s}\int d\mathbf{r}\varphi_j^{\prime\dag}(\mathbf{r})
\frac{\mathcal{V}'(\mathbf{r},-i\boldsymbol{\partial})}{w^{(0)}_r+
\sqrt{\phi_t}m}\frac{\mathcal{V}'(\mathbf{r},-i\boldsymbol{\partial})}
{w^{(0)}_{r'}+\sqrt{\phi_t}m}\varphi'_{j'}(\mathbf{r}) &=\frac{w^{(0)}_r-
\sqrt{\phi_t}m}{w^{(0)}_r+\sqrt{\phi_t}m}\delta_{j,j'},\\
\label{ort_chi_2}
\frac{\phi_t}{\phi_s}\int d\mathbf{r}\chi_j^{\prime\dag}(\mathbf{r})
\frac{\mathcal{V}'(\mathbf{r},-i\boldsymbol{\partial})}{\tilde{w}^{(0)}_q+
\sqrt{\phi_t}m}\frac{\mathcal{V}'(\mathbf{r},-i\boldsymbol{\partial})}
{\tilde{w}^{(0)}_{q'}+\sqrt{\phi_t}m}\chi'_{j'}(\mathbf{r}) &=
\frac{\tilde{w}^{(0)}_q-\sqrt{\phi_t}m}{\tilde{w}^{(0)}_q+\sqrt{\phi_t}m}\delta_{j,j'}.
\end{align}
Finally, from these relations and from Eqs. (\ref{xi_0}), (\ref{eta_0}),
(\ref{xi_pr_0}) and (\ref{eta_pr_0}) we derive immediately the equalities
\begin{align}
\label{ort_phi_3}
\int d\mathbf{r} \sqrt{\phi_s^3}u^{\dag}_j(\mathbf{r})\beta u_{j'}(\mathbf{r}) &=
\int d\mathbf{r} \sqrt{\phi_s^3}u^{\prime\dag}_j(\mathbf{r})\beta u'_{j'}(\mathbf{r})=
\frac{\sqrt{\phi_t}m}{w^{(0)}_r}\delta_{j,j'},\\
\label{ort_chi_3}
\int d\mathbf{r} \sqrt{\phi_s^3}v^{\dag}_j(\mathbf{r})\beta v_{j'}(\mathbf{r})&=
\int d\mathbf{r} \sqrt{\phi_s^3}v^{\prime\dag}_j(\mathbf{r})\beta v'_{j'}(\mathbf{r})=
-\frac{\sqrt{\phi_t}m}{\tilde{w}^{(0)}_q}\delta_{j,j'}
\end{align}
where we used the fact that [see Eq. (\ref{op_rot})]
\begin{equation}
\mathcal{R}_x(\vartheta)\beta\mathcal{R}^{\dag}_x(\vartheta)=\beta
\end{equation}
By the way, since they are often used in the main text, we also remind the
following transformation properties of the matrices $\boldsymbol{\alpha}$
\begin{align}
\label{alpha_x_p}
\mathcal{R}_x(\vartheta)\alpha_x\mathcal{R}^{\dag}_x(\vartheta) &=\alpha_x,\\
\label{alpha_y_p}
\mathcal{R}_x(\vartheta)\alpha_y\mathcal{R}^{\dag}_x(\vartheta) &=\alpha_y
\cos\vartheta+\alpha_z\sin\vartheta,\\
\label{alpha_z_p}
\mathcal{R}_x(\vartheta)\alpha_z\mathcal{R}^{\dag}_x(\vartheta) &=-\alpha_y
\sin\vartheta+\alpha_z\cos\vartheta
\end{align}
and of the position operator $\mathbf{r}$
\begin{align}
\label{x_p}
\mathcal{R}_x(\vartheta)x\mathcal{R}^{\dag}_x(\vartheta) &=x,\\
\label{y_p}
\mathcal{R}_x(\vartheta)y\mathcal{R}^{\dag}_x(\vartheta) &=y\cos\vartheta+z\sin\vartheta,\\
\mathcal{R}_x(\vartheta)z\mathcal{R}^{\dag}_x(\vartheta) &=-y\sin\vartheta+z\cos\vartheta.
\end{align}

\clearpage
%
%
\section*{Appendix B}
\setcounter{equation}{0}
\renewcommand{\theequation}{B\arabic{equation}}

We want to show explicitly here that
\begin{align}
\label{m_e_v_1}
\mathcal{I}_{r_-,n_g;r_+,n'_g} &\equiv \int d\mathbf{r}\sqrt{\phi_s^3}
u^{\dag}_{r_-,n_g}(\mathbf{r})\mathcal{I}(\mathbf{r},-i\boldsymbol{\partial})
u_{r_+,n'_g}(\mathbf{r})=0,\\
\mathcal{I}_{r_+,n_g;r_-,n'_g} &\equiv \int d\mathbf{r}\sqrt{\phi_s^3}
u^{\dag}_{r_+,n_g}(\mathbf{r})\mathcal{I}(\mathbf{r},-i\boldsymbol{\partial})
u_{r_-,n'_g}(\mathbf{r})=0
\end{align}
where $r_-=\{n_d+1,k,-1\}$ and $r_+=\{n_d,k,+1\}$. We will prove only the first
of these equalities being the other analogous. From Eq. (\ref{U_1p_ti}) we have
\begin{equation}
\mathcal{I}_{r_-,n_g;r_+,n'_g}=\int d\mathbf{r}\sqrt{\phi_s^3}
u^{\prime\dag}_{r_-,n_g}(\mathbf{r})\left[\sqrt{\phi_t}(g_P-g_M)\beta m-
(g_P-g_E)w^{(0)}_{r_-}\right]xu'_{r_+,n'_g}(\mathbf{r})
\end{equation}
where we used Eqs. (\ref{alpha_x_p}) and (\ref{x_p}) and the fact that
$w^{(0)}_{r_-}=w^{(0)}_{r_+}$. This equation can be rewritten in the form
[see Eq. (\ref{xi_pr_0})]
\begin{equation}
\label{m_e_v_2}
\begin{split}
\mathcal{I}_{r_-,n_g;r_+,n'_g}=\frac{w^{(0)}_{r_-}+\sqrt{\phi_t}m}
{2w^{(0)}_{r_-}}&\bigg\{\left[\sqrt{\phi_t}(g_P-g_M) m-(g_P-g_E)w^{(0)}_{r_-}\right]
\int d\mathbf{r} \varphi^{\prime\dag}_{r_-,n_g}(\mathbf{r})x\varphi'_{r_+,n'_g}(\mathbf{r})+\\
&+\left[-\sqrt{\phi_t}(g_P-g_M) m-(g_P-g_E)w^{(0)}_{r_-}\right]\times\\
&\times\frac{\phi_t}{\phi_s}\int d\mathbf{r}
\varphi^{\prime\dag}_{r_-,n_g}(\mathbf{r})
\frac{\mathcal{V}'(\mathbf{r},-i\boldsymbol{\partial})}{w^{(0)}_{r_-}+
\sqrt{\phi_t}m}x\frac{\mathcal{V}'(\mathbf{r},-i\boldsymbol{\partial})}
{w^{(0)}_{r_-}+\sqrt{\phi_t}m}\varphi'_{r_+,n'_g}(\mathbf{r})\bigg\}.
\end{split}
\end{equation}
Now, by observing that the spin of the two states is different we realize
that the first integral vanishes because of the orthonormalization relation
(\ref{ort_phi_1}). Also, from the definition (\ref{V_p_d}) of the operator
$\mathcal{V}'(\mathbf{r},-i\boldsymbol{\partial})$ and from its expression
(\ref{V_p}) we have that
\begin{equation}
\label{Vp_x_Vp}
\begin{split}
\mathcal{V}'(\mathbf{r},-i\boldsymbol{\partial})x\mathcal{V}'
(\mathbf{r},-i\boldsymbol{\partial}) &=[\mathcal{V}'(\mathbf{r},-i\boldsymbol{\partial}),x]
\mathcal{V}'(\mathbf{r},-i\boldsymbol{\partial})+x\mathcal{V}^{\prime 2}(\mathbf{r},
-i\boldsymbol{\partial})=\\
&=-i\sigma_x\mathcal{V}'(\mathbf{r},-i\boldsymbol{\partial})+x\mathcal{V}^{\prime 2}
(\mathbf{r},-i\boldsymbol{\partial})=\\
&=\sqrt{2eB}(a_d^{\dag}\sigma_x\sigma_--a_d\sigma_x\sigma_+)-\sigma_x\sigma_z
\frac{\partial}{\partial z}+\\
&+x\left[-\frac{\partial^2}{\partial z^2}+eB\left(2a_d^{\dag}a_d+1+\sigma_z\right)\right]
\end{split}
\end{equation}
where we used Eq. (\ref{V_p_2}). The only term of this operator that can change
the spin of the electron is the second one, but it can not change the value of
the quantum number $n_d$. This means that the second integral in Eq. (\ref{m_e_v_2})
also vanishes and this completes the prove of Eq. (\ref{m_e_v_1}).
\clearpage
%
%
\section*{Appendix C}
\setcounter{equation}{0}
\renewcommand{\theequation}{C\arabic{equation}}

We want to give here the explicit expression of the zero-order transverse ground
states and of the zero-order states corresponding to the first-excited Landau
levels both of the electron and of the positron. We remind that we used them to
calculate the transition matrix elements (\ref{M_E_cr_rot_2}),(\ref{U_p_m}),
(\ref{U_p_p}), (\ref{M_-}) and (\ref{M_+}). These states can be easily obtained
by substituting Eq. (\ref{V_p}) in Eqs. (\ref{U^P0_f}) and (\ref{V^P0_f}). We
give only the final results for the rotated states because, actually, we used
only them
\begin{align}
\label{U_gr_0}
U^{\prime (0)}_{0,k,-1,x_0}(\mathbf{r}) &=\frac{1}{\sqrt[4]{\phi_s^3}}
\sqrt{\frac{\varepsilon^{(0)}_k+\sqrt{\phi_t}m}{2\varepsilon^{(0)}_k}}
\left(
\begin{array}{c}
0\\
\Theta'_{0,x_0}(x,y)\\
-\sqrt{\frac{\phi_t}{\phi_s}}\frac{1}{\varepsilon^{(0)}_k+\sqrt{\phi_t}m}\left(\begin{array}{c}
0\\
k\Theta'_{0,x_0}(x,y)
\end{array}\right)
\end{array}\right)\frac{\exp(ikz)}{\sqrt{Z}},\\
\label{V_gr_0}
V^{\prime (0)}_{0,k,+1,x_0}(\mathbf{r}) &=\frac{1}{\sqrt[4]{\phi_s^3}}
\sqrt{\frac{\varepsilon^{(0)}_k+\sqrt{\phi_t}m}{2\varepsilon^{(0)}_k}}
\left(
\begin{array}{c}
-\sqrt{\frac{\phi_t}{\phi_s}}\frac{1}{\varepsilon^{(0)}_k+\sqrt{\phi_t}m}
\left(\begin{array}{c}
0\\
k\Theta'_{0,x_0}(x,y)
\end{array}\right)\\
0\\
\Theta'_{0,x_0}(x,y)
\end{array}\right)\frac{\exp(-ikz)}{\sqrt{Z}},\\
\label{U_pr_0_1}
U^{\prime (0)}_{0,k,+1,x_0}(\mathbf{r}) &=\frac{1}{\sqrt[4]{\phi_s^3}}
\sqrt{\frac{\mathcal{E}^{(0)}_k+\sqrt{\phi_t}m}{2\mathcal{E}^{(0)}_k}}
\left(
\begin{array}{c}
\Theta'_{0,x_0}(x,y)\\
0\\
\sqrt{\frac{\phi_t}{\phi_s}}\frac{1}{\mathcal{E}^{(0)}_k+\sqrt{\phi_t}m}
\left(\begin{array}{c}
k\Theta'_{0,x_0}(x,y)\\
i\sqrt{2eB}\Theta'_{1,x_0}(x,y)
\end{array}\right)
\\
\end{array}\right)\frac{\exp(ikz)}{\sqrt{Z}},\\
\label{U_pr_0_2}
U^{\prime (0)}_{1,k,-1,x_0}(\mathbf{r}) &=\frac{1}{\sqrt[4]{\phi_s^3}}
\sqrt{\frac{\mathcal{E}^{(0)}_k+\sqrt{\phi_t}m}{2\mathcal{E}^{(0)}_k}}
\left(
\begin{array}{c}
0\\
\Theta'_{1,x_0}(x,y)\\
-\sqrt{\frac{\phi_t}{\phi_s}}\frac{1}{\mathcal{E}^{(0)}_k+\sqrt{\phi_t}m}
\left(\begin{array}{c}
i\sqrt{2eB}\Theta'_{0,x_0}(x,y)\\
k\Theta'_{1,x_0}(x,y)
\end{array}\right)
\\
\end{array}\right)\frac{\exp(ikz)}{\sqrt{Z}},\\
\label{V_pr_0_1}
V^{\prime (0)}_{0,k,-1,x_0}(\mathbf{r}) &=\frac{1}{\sqrt[4]{\phi_s^3}}
\sqrt{\frac{\mathcal{E}^{(0)}_k+\sqrt{\phi_t}m}{2\mathcal{E}^{(0)}_k}}
\left(
\begin{array}{c}
\sqrt{\frac{\phi_t}{\phi_s}}\frac{1}{\mathcal{E}^{(0)}_k+\sqrt{\phi_t}m}
\left(\begin{array}{c}
-k\Theta'_{0,x_0}(x,y)\\
i\sqrt{2eB}\Theta'_{1,x_0}(x,y)
\end{array}\right)
\\
-\Theta'_{0,x_0}(x,y)\\
0
\end{array}\right)\frac{\exp(-ikz)}{\sqrt{Z}},\\
\label{V_pr_0_2}
V^{\prime (0)}_{1,k,+1,x_0}(\mathbf{r}) &=\frac{1}{\sqrt[4]{\phi_s^3}}
\sqrt{\frac{\mathcal{E}^{(0)}_k+\sqrt{\phi_t}m}{2\mathcal{E}^{(0)}_k}}
\left(
\begin{array}{c}
\sqrt{\frac{\phi_t}{\phi_s}}\frac{1}{\mathcal{E}^{(0)}_k+\sqrt{\phi_t}m}
\left(\begin{array}{c}
i\sqrt{2eB}\Theta'_{0,x_0}(x,y)\\
-k\Theta'_{1,x_0}(x,y)
\end{array}\right)\\
0\\
\Theta'_{1,x_0}(x,y)
\end{array}\right)\frac{\exp(-ikz)}{\sqrt{Z}}.
\end{align}
In these expressions we have used the definitions (\ref{Phi}) and (\ref{Chi}) for
the two-dimensional spinors $\Phi'_J(\mathbf{r})$ and $\mathrm{X}'_J(\mathbf{r})$
and the definitions (\ref{E_0}) and (\ref{E_1}) for the energies $\varepsilon^{(0)}_k$
and $\mathcal{E}^{(0)}_k$.
\clearpage
%
%
\section*{Appendix D}
\setcounter{equation}{0}
\renewcommand{\theequation}{D\arabic{equation}}

In this appendix we will give an example which explains how to calculate the matrix
elements (\ref{U_p_m}), (\ref{U_p_p}) and (\ref{M_-}) and (\ref{M_+}). In particular,
we will calculate the integral
\begin{equation}
\begin{split}
\left\langle U^{\prime(1)}_{J_0}\big\vert\alpha_-\big\vert U^{\prime(1)}_{J_0}\right\rangle
&=\int d\mathbf{r}\sqrt{\phi_s^3}U^{\prime(0)\dag}_{0,k,-1,x_0}(\mathbf{r})\alpha_-
\left[\mathsf{B}^{(1)}_{k,x_0}U^{(0)}_{1,k,-1,x_0}(\mathbf{r})-
i\mathsf{C}^{(1)}_{k,x_0}U^{(0)}_{0,k,+1,x_0}(\mathbf{r})+\right.\\
&\qquad\qquad\qquad\qquad\qquad\qquad\left.+
i\mathsf{E}^{(1)}_{k,x_0}V^{(0)\dag}_{0,-k,-1,x_0}(\mathbf{r})-
\mathsf{F}^{(1)}_{k,x_0}V^{(0)}_{1,-k,+1,x_0}(\mathbf{r})\right].
\end{split}
\end{equation}
We first observe that, given a generic spinor $S$,
\begin{equation}
S=\left(
\begin{array}{c}
S_1\\
S_2\\
S_3\\
S_4
\end{array}\right)
\end{equation}
then
\begin{equation}
\alpha_-S=\left(
\begin{array}{cccc}
0 & 0 & 0 & 0\\
0 & 0 & 1 & 0\\
0 & 0 & 0 & 0\\
1 & 0 & 0 & 0
\end{array}\right)
\left(
\begin{array}{c}
S_1\\
S_2\\
S_3\\
S_4
\end{array}\right)=\left(
\begin{array}{c}
0\\
S_3\\
0\\
S_1
\end{array}\right).
\end{equation}

In this way, from the expressions (\ref{U_pr_0_1})-(\ref{V_pr_0_2}) of the zero-order
electron and positron eigenstates we see that
\begin{equation}
\begin{split}
&\alpha_-\left[\mathsf{B}^{(1)}_{k,x_0}U^{(0)}_{1,k,-1,x_0}
(\mathbf{r})-i\mathsf{C}^{(1)}_{k,x_0}U^{(0)}_{0,k,+1,x_0}(\mathbf{r})+
i\mathsf{E}^{(1)}_{k,x_0}V^{(0)\dag}_{0,-k,-1,x_0}(\mathbf{r})-
\mathsf{F}^{(1)}_{k,x_0}V^{(0)}_{1,-k,+1,x_0}(\mathbf{r})\right]=\\
&=-\frac{i}{\sqrt[4]{\phi_s^3}}\sqrt{\frac{\mathcal{E}^{(0)}_k+\sqrt{\phi_t}m}
{2\mathcal{E}^{(0)}_k}}
\left(
\begin{array}{c}
0\\
\sqrt{\frac{\phi_t}{\phi_s}}\frac{\sqrt{2eB}\mathsf{B}^{(1)}_{k,x_0}+
k\mathsf{C}^{(1)}_{k,x_0}}{\mathcal{E}^{(0)}_k+\sqrt{\phi_t}m}+\mathsf{E}^{(1)}_{k,x_0}\\
0\\
\mathsf{C}^{(1)}_{k,x_0}-\sqrt{\frac{\phi_t}{\phi_s}}\frac{k\mathsf{E}^{(1)}_{k,x_0}-
\sqrt{2eB}\mathsf{F}^{(1)}_{k,x_0}}{\mathcal{E}^{(0)}_k+\sqrt{\phi_t}m}
\end{array}\right)\Theta'_{0,x_0}(x,y)\frac{\exp(ikz)}{\sqrt{Z}}.
\end{split}
\end{equation}
By using the expression (\ref{U_gr_0}) for $U^{\prime(0)}_{0,k,-1,x_0}(\mathbf{r})$ we obtain
\begin{equation}
\begin{split}
\left\langle U^{\prime(1)}_{J_0}\big\vert\alpha_-\big\vert U^{\prime(1)}_{J_0}\right\rangle
&=-i\sqrt{\frac{\varepsilon^{(0)}_k+\sqrt{\phi_t}m}{2\varepsilon^{(0)}_k}}
\sqrt{\frac{\mathcal{E}^{(0)}_k+\sqrt{\phi_t}m}{2\mathcal{E}^{(0)}_k}}\times\\
&\times\left[\sqrt{\frac{\phi_t}{\phi_s}}\frac{\sqrt{2eB}\mathsf{B}^{(1)}_{k,x_0}+
k\mathsf{C}^{(1)}_{k,x_0}}{\mathcal{E}^{(0)}_k+\sqrt{\phi_t}m}+\mathsf{E}^{(1)}_{k,x_0}-
\right.\\
&\left.-\sqrt{\frac{\phi_t}{\phi_s}}\frac{k}{\varepsilon^{(0)}_k+\sqrt{\phi_t}m}
\left(\mathsf{C}^{(1)}_{k,x_0}-\sqrt{\frac{\phi_t}{\phi_s}}\frac{k\mathsf{E}^{(1)}_{k,x_0}-
\sqrt{2eB}\mathsf{F}^{(1)}_{k,x_0}}{\mathcal{E}^{(0)}_k+\sqrt{\phi_t}m}\right)\right].
\end{split}
\end{equation}
At this point we have to substitute the coefficients $\mathsf{B}^{(1)}_{k,x_0}$,
$\mathsf{C}^{(1)}_{k,x_0}$, $\mathsf{E}^{(1)}_{k,x_0}$ and $\mathsf{F}^{(1)}_{k,x_0}$
by means of Eqs. (\ref{B_c}), (\ref{C_c}), (\ref{E_c}) and (\ref{F_c}) and many terms
cancel each other. After some algebra we obtain
\begin{equation}
\begin{split}
\left\langle U^{\prime(1)}_{J_0}\big\vert\alpha_-\big\vert U^{\prime(1)}_{J_0}
\right\rangle &=-i\frac{\varepsilon^{(0)}_k+\sqrt{\phi_t}m}{2\varepsilon^{(0)}_k}
\frac{\mathcal{E}^{(0)}_k+\sqrt{\phi_t}m}{2\mathcal{E}^{(0)}_k}\times\\
&\times\left\{\frac{1}{\mathcal{E}^{(0)}_k-\varepsilon^{(0)}_k}
\left[\sqrt{\phi_t}(g_M-g_P)m+\frac{g_P}{2}\big(\mathcal{E}^{(0)}_k+
\varepsilon^{(0)}_k\big)-g_E\varepsilon^{(0)}_k\right]\times\right.\\
&\qquad\times\sqrt{\frac{\phi_t}{\phi_s}}\frac{1}{\mathcal{E}^{(0)}_k+\sqrt{\phi_t}m}+\\
&\qquad+\frac{1}{\mathcal{E}^{(0)}_k+\varepsilon^{(0)}_k}\left[\sqrt{\phi_t}(g_M-g_P)m-
\frac{g_P}{2}\big(\mathcal{E}^{(0)}_k-\varepsilon^{(0)}_k\big)-
g_E\varepsilon^{(0)}_k\right]\times\\
&\qquad\times\sqrt{\frac{\phi_t}{\phi_s}}\frac{1}{\mathcal{E}^{(0)}_k+\sqrt{\phi_t}m}-\\
&\qquad-\frac{1}{\mathcal{E}^{(0)}_k-\varepsilon^{(0)}_k}\left[\sqrt{\phi_t}(g_M-g_P)m-
\frac{g_P}{2}\big(\mathcal{E}^{(0)}_k+\varepsilon^{(0)}_k\big)+g_E\varepsilon^{(0)}_k\right]
\times\\
&\qquad\times\sqrt{\frac{\phi_t}{\phi_s}}\frac{k}{\mathcal{E}^{(0)}_k+\sqrt{\phi_t}m}
\frac{\phi_t}{\phi_s}\frac{k}{\big(\mathcal{E}^{(0)}_k+\sqrt{\phi_t}m\big)
\big(\varepsilon^{(0)}_k+\sqrt{\phi_t}m\big)}-\\
&\qquad-\frac{1}{\mathcal{E}^{(0)}_k+\varepsilon^{(0)}_k}\left[\sqrt{\phi_t}
(g_M-g_P)m+\frac{g_P}{2}\big(\mathcal{E}^{(0)}_k-\varepsilon^{(0)}_k\big)+
g_E\varepsilon^{(0)}_k\right]\times\\
&\qquad\left.\times\sqrt{\frac{\phi_t}{\phi_s}}\frac{k}{\mathcal{E}^{(0)}_k+
\sqrt{\phi_t}m}\frac{\phi_t}{\phi_s}\frac{k}{\big(\mathcal{E}^{(0)}_k+
\sqrt{\phi_t}m\big)\big(\varepsilon^{(0)}_k+\sqrt{\phi_t}m\big)}\right\}.
\end{split}
\end{equation}
Finally, if we use the equalities
\begin{align}
1-\frac{\phi_t}{\phi_s}\frac{k^2}{\big(\varepsilon^{(0)}_k+\sqrt{\phi_t}m\big)^2}
&=\frac{2\sqrt{\phi_t}m}{\varepsilon^{(0)}_k+\sqrt{\phi_t}m},\\
1+\frac{\phi_t}{\phi_s}\frac{k^2}{\big(\varepsilon^{(0)}_k+\sqrt{\phi_t}m\big)^2}
&=\frac{2\varepsilon^{(0)}_k}{\varepsilon^{(0)}_k+\sqrt{\phi_t}m}
\end{align}
we have
\begin{equation}
\begin{split}
\left\langle U^{\prime(1)}_{J_0}\big\vert\alpha_-\big\vert U^{\prime(1)}_{J_0}
\right\rangle &=-i\sqrt{\frac{\phi_t}{\phi_s}}\frac{1}{2\varepsilon^{(0)}_k
\mathcal{E}^{(0)}_k}\times\\
&\times\left\{\frac{1}{\mathcal{E}^{(0)}_k-\varepsilon^{(0)}_k}\times\right.\\
&\qquad\times\bigg[\sqrt{\phi_t}m(g_M-g_P)\sqrt{\phi_t}m+\left[\frac{g_P}{2}
\big(\mathcal{E}^{(0)}_k+\varepsilon^{(0)}_k\big)-g_E\varepsilon^{(0)}_k\right]
\varepsilon^{(0)}_k\bigg]+\\
&\qquad+\frac{1}{\mathcal{E}^{(0)}_k+\varepsilon^{(0)}_k}\times\\
&\qquad\left.\times\bigg[\sqrt{\phi_t}m(g_M-g_P)\sqrt{\phi_t}m-\left[\frac{g_P}{2}
\big(\mathcal{E}^{(0)}_k-\varepsilon^{(0)}_k\big)+g_E\varepsilon^{(0)}_k\right]
\varepsilon^{(0)}_k\bigg]\right\}=\\
&=i\frac{1}{2eB}\sqrt{\frac{\phi_s}{\phi_t}}\left[(g_M-g_P) \frac{\phi_tm^2}
{\varepsilon^{(0)}_k}+(g_P-g_E)\varepsilon^{(0)}_k\right]=\\
&=i\frac{\sqrt{\phi_sm^2+k^2}}{2eB}\left(\frac{g_t}{\phi_t}+\frac{g_s}{\phi_s}
\frac{k^2}{\phi_sm^2+k^2}\right)
\end{split}
\end{equation}
that is one of the equalities in Eq. (\ref{M_E_U_V}). We say again that all
the other matrix elements can be calculated with the same technique.
\clearpage
%
%

\clearpage
\end{document}